\documentclass[a4paper,11pt]{article}
\pdfoutput=1
\usepackage{jheppub}
\usepackage{amsmath}
\usepackage{amsthm}
\usepackage{amsfonts}
\usepackage{amssymb}
\usepackage{textcomp}
\usepackage{graphicx}
\usepackage{bm}
\usepackage{color}
\usepackage{verbatim}
\usepackage{subcaption}
\usepackage{graphicx,color}
\usepackage{epsfig}
\usepackage{slashed}
\usepackage{bbold}
\usepackage{soul}

\definecolor{orange}{rgb}{1,0.5,0}
\definecolor{col1}{RGB}{153, 52, 121}
\definecolor{dgreen}{rgb}{0,0.55,0}
\definecolor{pink}{rgb}{1,0.08,0.58}

\usepackage{amsfonts,amsmath,amssymb}

\newcommand{\p}{\partial}

\newcommand{\la}{\langle}
\newcommand{\ra}{\rangle}

\newcommand{\rar}{\rightarrow}

\newcommand{\ua}{\uparrow}
\newcommand{\da}{\downarrow}

\theoremstyle{definition}

 \title{
 \Huge Isolated zeros destroy Fermi surface 
 \\
 in holographic models with a lattice \color{black}}
 
 \author{Floris Balm, \footnote{https://orcid.org/0000-0001-9387-3203}}
 \author{Alexander Krikun,\footnote{https://orcid.org/0000-0001-8789-8703}}
 \author{Aurelio Romero-Berm\'udez,}
 \author{\\Koenraad Schalm and}
 \author{Jan Zaanen}

\affiliation{Instituut-Lorentz for Theoretical Physics, $\Delta ITP$, Leiden University, Niels Bohrweg 2, Leiden 2333CA, The Netherlands}

\emailAdd{balm@lorentz.leidenuniv.nl}
\emailAdd{krikun@lorentz.leidenuniv.nl}
\emailAdd{romero@lorentz.leidenuniv.nl}
\emailAdd{kschalm@lorentz.leidenuniv.nl}
\emailAdd{jan@lorentz.leidenuniv.nl}

\abstract{
We study the fermionic spectral density in a strongly correlated quantum system described by a gravity dual. In the presence of periodically modulated chemical potential, which models the effect of the ionic lattice, we explore the shapes of the corresponding Fermi surfaces, defined by the location of peaks in the spectral density at the Fermi level. 
We find that at strong lattice potentials sectors of the Fermi surface are unexpectedly destroyed and the Fermi surface becomes an arc-like disconnected manifold.
We explain this phenomenon in terms of a collision of the Fermi surface pole with zeros of the fermionic Green's function, which are explicitly computable in the holographic dual.
}

\begin{document}

\maketitle

\section{\label{sec:Intro}Introduction}

The physics of strongly correlated electron systems remains a major puzzle in modern condensed matter theory. The possible deviations from conventional Fermi liquid behavior are simultaneously extremely interesting and extremely hard to study. Nonetheless, evidence coming from experiments in high temperature superconductors and other strange metallic systems points out that such non-Fermi liquid systems do exist in nature and  display  many unconventional phenomena. The defining feature of non-Fermi liquid behavior is the absence of long lived quasiparticles anchored on a well-defined Fermi surface, which could be used as building blocks for Fermi liquid perturbation theory. Such signatures of a destruction of the quasiparticles are seen in the angle resolved photoemission (ARPES) studies of experimentally realized strange metals \cite{loeser1996excitation}. In high T$_c$ superconductors in the normal phase, the spectral width, or the inverse lifetime, of the fermionic excitation at the Fermi level becomes unnaturally broad in the anti-nodal directions, whereas one still observes well defined quasiparticles at the nodes \cite{shen2005nodal,zhou2004dichotomy} -- the nodal-antinodal dichotomy. In the pseudogap phase the phenomenon of so-called Fermi arcs is even more striking: the sharp Fermi surface simply ends at a point in the Brillouin zone, which is topologically forbidden for a Fermi liquid-like system \cite{yoshida2006systematic}. A careful theoretical understanding of these phenomena has been hampered by an absence of conceptually new approaches that do not rely on a stable quasiparticle description.

The holographic duality provides such a conceptually novel way to treat the strongly correlated systems without the need to postulate the quasiparticle description to begin with; for a review see \cite{Zaanen:2015oix}. It has been shown in the earlier works \cite{Cubrovic:2009ye,Faulkner:2009wj,Faulkner:2011tm} that one can obtain a spectrum with or without the long lived fermionic quasiparticles depending on the parameters of the holographic model. The natural question arises of whether it is possible to achieve the transition between these regimes as a function of direction in the Brillouin zone, as it is observed in real materials. This is the question we address in the present work. Most of the earlier works in holography on single fermion spectral functions are restricted to isotropic setups, with the rare exceptions including \cite{Liu:2012tr,cremonini2018holographic,Ling:2013aya,Bagrov:2016cnr,Ling:2014bda,Cremonini:2019fzz}. In order to study anisotropy and the effects of the Brillouin zone boundary, we introduce a  periodic modulation of the chemical potential, which mimics the ionic lattice, breaks the rotational symmetry down to a discrete group and introduces Bloch momenta.\footnote{For the sake of technical simplicity we focus on a quasi-one-dimensional lattice, i.e. we only consider the modulation in one spatial direction.} 
A similar study has been performed recently in \cite{cremonini2018holographic,Cremonini:2019fzz} which also includes a spontaneous breaking of translational invariance. In our case we restrict ourselves 
to a simpler setup that includes only an explicit periodic lattice, without spontaneous striped order. This allows us to perform the analysis of our results in the cleanest possible way. Similarly, in order to isolate the effects of the periodic lattice, we don't consider any non-minimal interaction terms in the Lagrangian of fermions coupled to gravity in holographic dual description.

As expected, we observe the generic effects of a periodic potential: the fermionic dispersion relation becomes multivalued in the first Brillouin zone due to the presence of lattice copies from the neighbouring zones.  
Umklapp gaps appear from the interaction between these copies and results in the formation of Fermi pockets. These basic effects had already been observed before in various holographic models with periodic potentials  \cite{Liu:2012tr,Ling:2013aya,cremonini2018holographic}. It simply shows the universality of  Umklapp at the boundary of the Brillouin zone in fermionic responses in periodic potentials.
However, we report here that, for  substantially strong lattice potentials (much stronger then considered previously in i.e. \cite{Liu:2012tr,Ling:2013aya}), a \textbf{novel physical effect} appears: the partial destruction of the Fermi surface due to the interaction of poles and zeros in the Green's function. 
A noticable spectral weight suppression in strong holographic mixed spontaneous and explicit lattices was observed earlier in \cite{cremonini2018holographic,Cremonini:2019fzz}.
However the effect which we observe is different:
in our purposely simple tractable model we can completely identify that there is not just suppression, but that the Fermi surface is actually destroyed due to a collision of the defining pole with a zero in the Green's function.\footnote{For a discussion on the possible origins of the anisotropic spectral weight suppression observed in \cite{cremonini2018holographic,Cremonini:2019fzz} see \cite{bagrovLattice}}.

Zeros of the fermionic Green's function have been observed in holographic models in several contexts. 
One kind of zeros, the ``alternative quantization zeros'', has been pointed out in the early works using bottom-up models  \cite{Cubrovic:2009ye,Faulkner:2009wj,Faulkner:2011tm} as well as in top-down constructions \cite{dewolfe2015fermi}. The existence of these zeros is understood straightforwardly within the holographic approach: they originate from the fact that, for a range of the parameters, a particular holographic theory can be treated as a dual to two distinct quantum theories on the boundary \cite{Klebanov:1999tb}. These two treatments are the ``direct'' and ``alternate'' quantization of the boundary operators and the simultaneous existence of both leads to the appearance of zeros in the Green's function of one theory, precisely at the point where the other one has poles. 
As we will see below, in the presence of a strong lattice potential these ``alternative quantization zeros'' approach the pole corresponding to the putative Fermi surface.
This proximity kills the peak in the spectral response.
The origin of this type of zeros is quite clear from the holographic point of view and universal in that context, but their interpretation in terms of conventional condensed matter theory remains elusive. We shall comment on possible interpretations in the discussion section. 
The phenomenon we observe is somewhat similar to the ``pole-skipping'' in holographic correlators for hydrodynamic energy-density modes, discussed recently in \cite{Grozdanov:2017ajz,Blake:2017ris,Blake:2018leo,Grozdanov:2018kkt,Blake:2019otz,Grozdanov:2018kkt} and for fermionic modes in 
\cite{Ceplak:2019ymw}. There one also observes the line of poles in the spectral density being cut by the line of zeros.

This holographic understanding of controlled zeros in the Green's function has already been exploited in attempts to describe the zeros in the spectrum arising from Mott physics \cite{edalati2011dynamical,vanacore2014minding, alsup2014duality,vanacore2018evolution,Seo:2018hrc,Chakrabarti:2019gow}. One mechanism relies on an extra Pauli or dipole coupling present in the fermionic Lagrangian. It has been shown that in the particular case of massless bulk fermions this can partially convert the Fermi surface into the line of zeros. We intentionally do not include the extra coupling in our model, which allows us to distinguish the phenomenon we observe here from the one mentioned in those works. 
On the other hand, zeros in spectral functions can have several origins but true Mott physics is intrinsically linked to the presence of a lattice and translational symmetry breaking as studied here.

The spectral signature of zeros colliding with poles/peaks is a very identifiable characteristic and for that reason of high interest.
More recently, a new zero-pole collision has been found  \cite{Gnezdilov:2018qdu} unrelated to holography. It was shown that in the presence of a quantum critical continuum coupled with two systems with discrete spectra, the spectrum of one such system has a characteristic zero at the resonance of the other. And this zero may collide with a pole. 
This effect is in the same class as the Fano resonance, where the spectrum of a continuum theory interacting with a discrete system has a zero at the resonance frequency of the latter.
Alongside with our main finding we observe this new class of ``resonant zeros'' in our holographic model. The reason is that the near horizon geometry generically encodes a certain type of the quantum critical continuum and the periodic potential gives rise to the many copies of the discrete particle dispersion spectra within the first Brillouin zone. The fermion spectral function precisely probes a discrete sector coupled via continuum to other discrete spectra. The effect of these zeros is also interesting, but not so spectacular as the one from ``alternative quantization'' ones. These discrete-continuum-discrete resonant zeros cut through the Fermi surface and destroy the quasiparticle peak at a particular point, but they do not remove  extended intervals from the Fermi surface.

Our finding that ``alternative quantization'' zeros interfere with Fermi-surface poles in holographic models with strong lattice potentials is theoretical and it is our thorough understanding of a peculiar aspect of holographic theories that allows us to unambiguously identify this mechanism. The result strikingly resembles the phenomenon of Fermi arcs, seen in the pseudogap. The creation of Fermi arcs was already the motivation for the holographic studies \cite{edalati2011dynamical,vanacore2014minding, alsup2014duality,vanacore2018evolution,Seo:2018hrc,Chakrabarti:2019gow}, but in our work they are directly tied to the presence of the lattice and anchored to the directional pattern of the Brillouin zone. Undoped Mott insulators are known to have zero responses in the single particle fermionic spectral function. 
In conventional condensed matter theory there are attempts to explain the formation of Fermi arcs in the pseudogap phase of the cuprates originating from these Mott zeros in the spectral density \cite{phillips2006mottness,stanescu2007theory,phillips2009mottness,imada2011theory,dave2013absence,Volovik_2018}. In these approaches, it is argued that, due to strong interactions, the self-energy of the quasiparticle diverges forcing the ``dressed'' Green's function to vanish at certain points in the phase space. These zeros do not violate the Luttinger count \cite{dzyaloshinskii2003some}, but render the Fermi surface disconnected. It would be interesting to determine whether there is any connection between our results and this other approach, but as we shall discuss in the conclusion, there are a number of fundamental open questions that will require significant further study.

The paper is organized as follows, in the first two sections we give an overview of basic features of ordinary Fermi surfaces in a periodic potential (Sec.\,\ref{sec:umklapp}) as well as the holographic fermionic response in the case of a simple isotropic background (Sec.\,\ref{sec:holographicFS}).
In Section \ref{sec:model}, we describe the model and the method. The main result is presented in Section \ref{sec:result} and we draw our conclusions in Section \ref{sec:conclusion}. The appendices are devoted to the details of our treatment in the homogeneous black hole background (App.\,\ref{app:RN_fermions}), the construction of periodic backgrounds (App.\,\ref{app:numerics}) and the analysis of the fermionic response (App.\,\ref{app:fermi_numerics}). We also explain the Brillouin zone representation of the Green's function in Appendix \ref{app:BZ}.

\section{\label{sec:umklapp}Umklapp scattering and Fermi pockets in unidirectional potential}

We start by recalling the basic features of the fermionic spectral function in a periodic potential. We wish to illustrate that at strong lattice potentials there are several non-linear responses, that are normally not considered in linear analysis of Bloch wave physics and Umklapp. Nevertheless these non-linear responses follow from a straightforward calculation. We consider a non-interacting Dirac fermion $\bar{\psi}$ in 2+1 dimensions in the presence of a periodically modulated chemical potential $\mu(x) = \mu [1 +\lambda\sin(p x)]$ along the $x$-direction only. In order to facilitate the analysis of our main results, we 
keep the $y$-direction homogeneous. The Dirac equation reads (here the bars denote a $2+1$ dimensional toy-model in order to avoid confusion with our main treatment below)
\begin{gather}
\label{eq:flat_Dirac}
\left[\bar{\gamma}^\mu \p_\mu - i \mu(x) \bar{\gamma}^t \right] \bar{\psi}(x,y,t) = 0, \\
\label{eq:mux}
\mu(x) \sim \mu(x + 2 n \pi/p) =  \mu \left(1 +\lambda\sin(p x) \right) \\
\notag
\bar{\gamma}^t = i \sigma_1, \qquad \bar{\gamma}^x =  \sigma_2, \qquad \bar{\gamma}^y = \sigma_3\,,
\end{gather}
where $\sigma_{i}$ are the Pauli matrices.
We introduce frequency,  momentum in $y$-direction as well as the Bloch momentum in $x$-direction
\begin{equation}
\bar{\psi}(x,y,t) \equiv e^{i (k_x x + k_y y - \omega t)} \bar{\psi}_k(x), \qquad  \bar{\psi}_k(x) \sim \bar{\psi}_k(x + 2 n \pi/p), n \in \mathbb{N},
\end{equation}
to get the equation on the Bloch wave function $\bar{\psi}_k(x)$, which is by construction periodic with the same period as the potential $\mu(x)$:  
\begin{gather}
\label{eq:toy_Dirac}
\left[- \sigma_1 (\omega - \mu(x))  + i \sigma_2 k_x + i \sigma_3 k_y + \sigma_2 \p_x  \right] \bar{\psi}_k(x) =0.
\end{gather}
Here, the Bloch wave function $\bar{\psi}_k = \{\bar{\psi}_k^\ua,\bar{\psi}_k^\da \} $ is a 2-component spinor and the Dirac operator is a $2\times2$ matrix differential operator. 
Since the Bloch wave function has the same period as the potential,  it is convenient to expand it in the Fourier series: 
\begin{equation}
\begin{pmatrix}
\bar{\psi}_k^\ua(x) \\ \bar{\psi}_k^\da(x)
\end{pmatrix} = \sum_l \begin{pmatrix}
{a_k^\ua}_l \\ {a_k^\da}_l
\end{pmatrix} e^{i l p x}.
\end{equation}
In this representation, the Dirac equation \eqref{eq:toy_Dirac} turns into a matrix equation for the vector wave function (let us drop the $k$ index for now) $\vec{a} = \{\dots, a^\ua_l, a^\da_l, a^\ua_{l+1}, a^\da_{l+1}\dots \}$  
\begin{gather}
\notag
M\cdot \vec{a} = \vec{0} \\
\label{equ:vector_toy_dirac}
M =
\begin{pmatrix}
\ddots		&						& \cdots 				& \cdots 			& 					&		\\
		& -(k_x - 2p) - \delta\omega	& -i k_y 				& - i\lambda\mu/2 	& 0					&		\\
\vdots	& i k_y 				& (k_x - 2 p) - \delta\omega	& 0					& - i\lambda\mu/2		& \vdots\\
		& i\lambda\mu/2 			& 0						& -(k_x - p) - \delta\omega&  - i k_y		  	&		\\
		& 0 					& -i \lambda\mu/2			& i k_y				& (k_x - p) - \delta\omega&		\\
		&  \cdots 				& \cdots 				& 					& 					&	\ddots				
\end{pmatrix},
\end{gather}
where $\delta \omega \equiv \omega - \mu$ for brevity.
The Green's function in the Fourier mode representation is simply the inverse of the Dirac operator: $G_{lm} = (M^{-1})_{lm}$. 
In ARPES experiments, the most important element of the Green's function is the $G_{00}$ component, which characterizes the projection of the fermionic linear response function in a crystal on the plane-wave states of the incident photon and photo-electron. Therefore, in what follows, we will be focusing on the spectral function associated with the $G_{00}$ component of the fermionic Green's function
\begin{equation}
\label{equ:spect_density_lattice}
A(\omega,k) = \mathrm{Im}\ \mathrm{Tr} G_{00}(\omega,k),
\end{equation}
where the trace is taken over the spin states. Throughout the paper we assume a practical definition of the Fermi surface as the locus of maxima of the spectral density in the momentum plane at the Fermi level.

\newcommand{\myWidL}{0.458}
\newcommand{\myWidR}{0.41}
\begin{figure}[t!]
\center
\begin{tabular}{ll}
\includegraphics[width=\myWidL\linewidth]{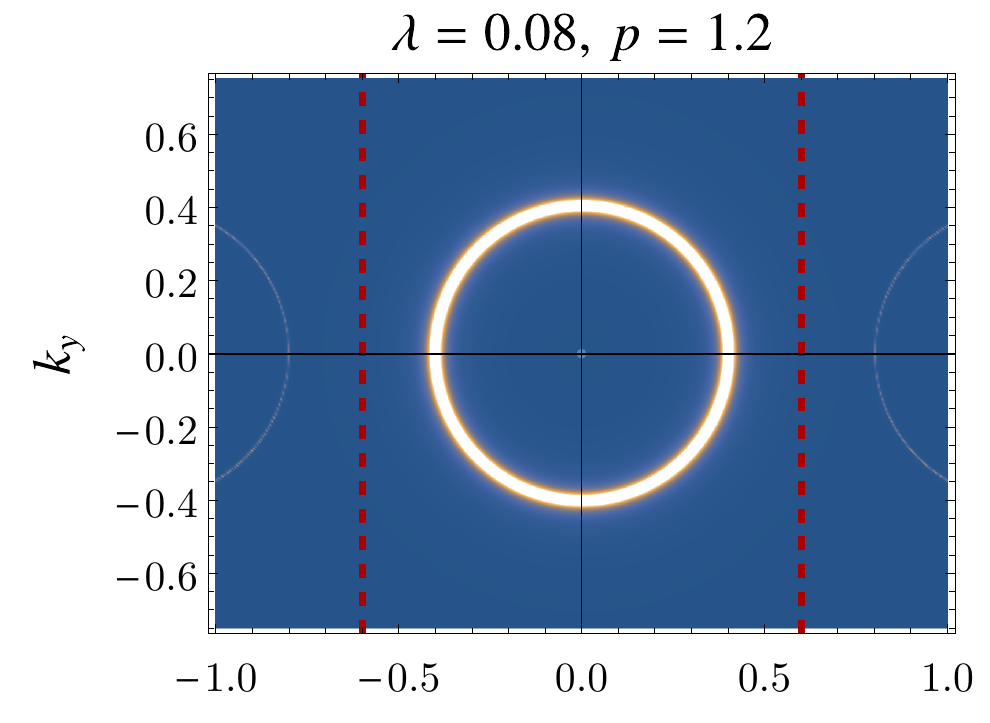} & \includegraphics[width=\myWidR\linewidth]{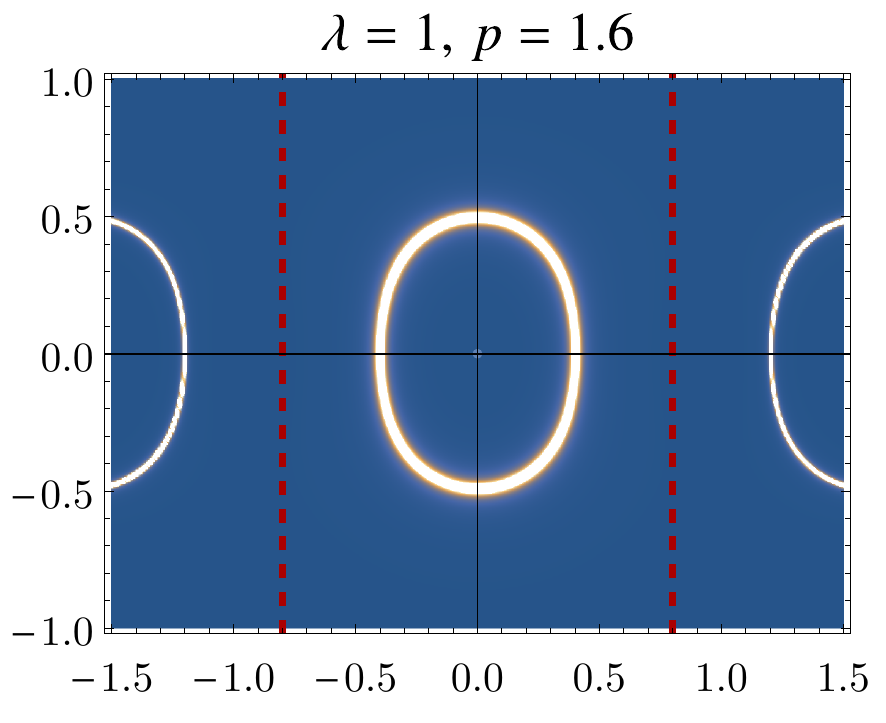} \\[-0.15cm]
\includegraphics[width=\myWidL\linewidth]{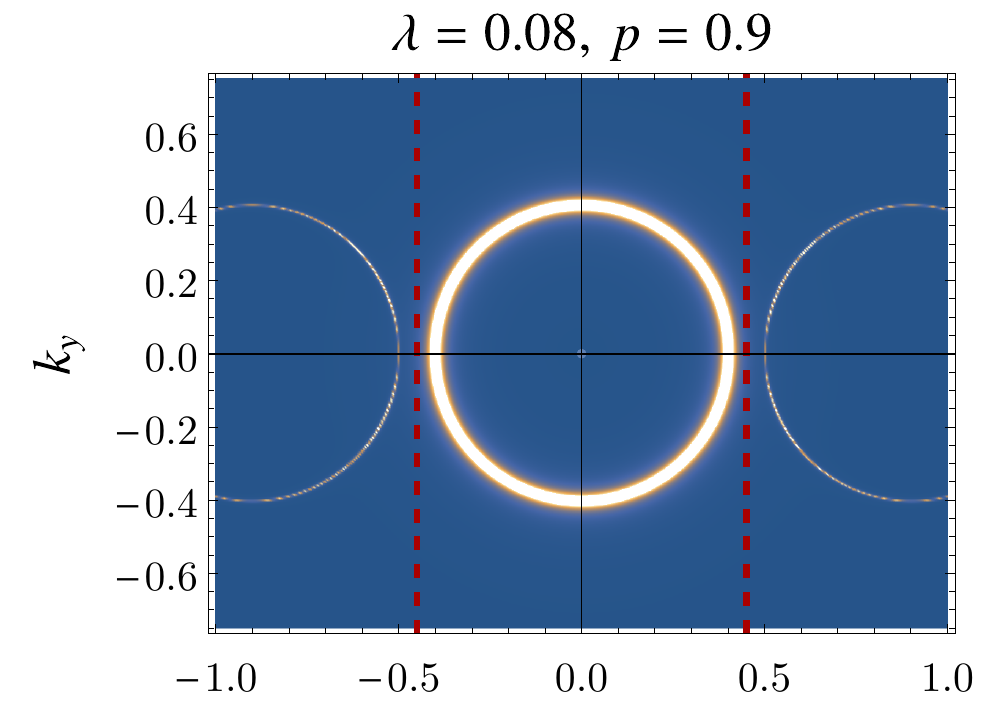} &\includegraphics[width=\myWidR\linewidth]{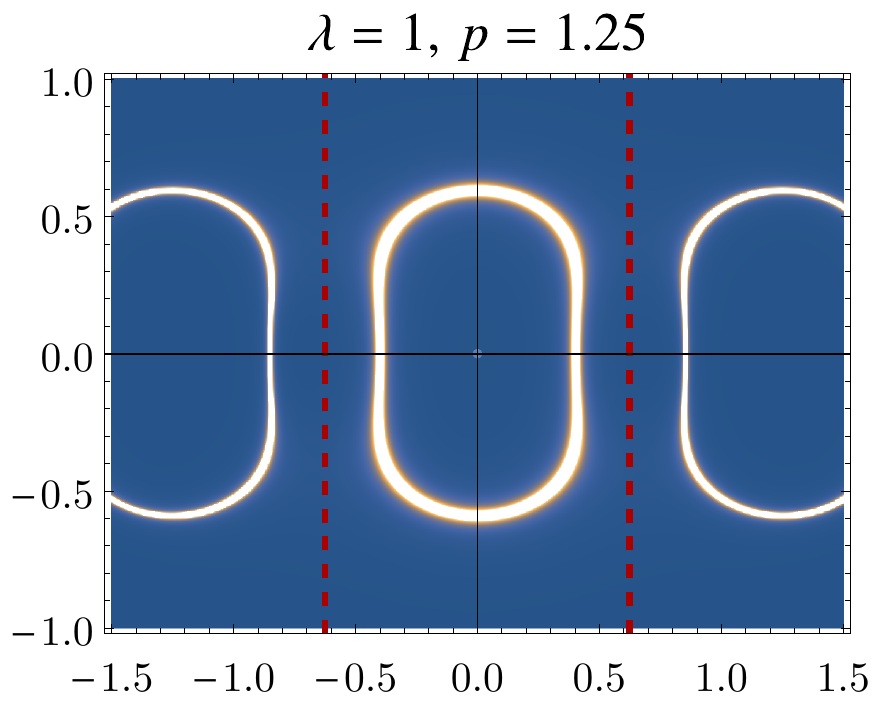} \\[-0.15cm]
\includegraphics[width=\myWidL\linewidth]{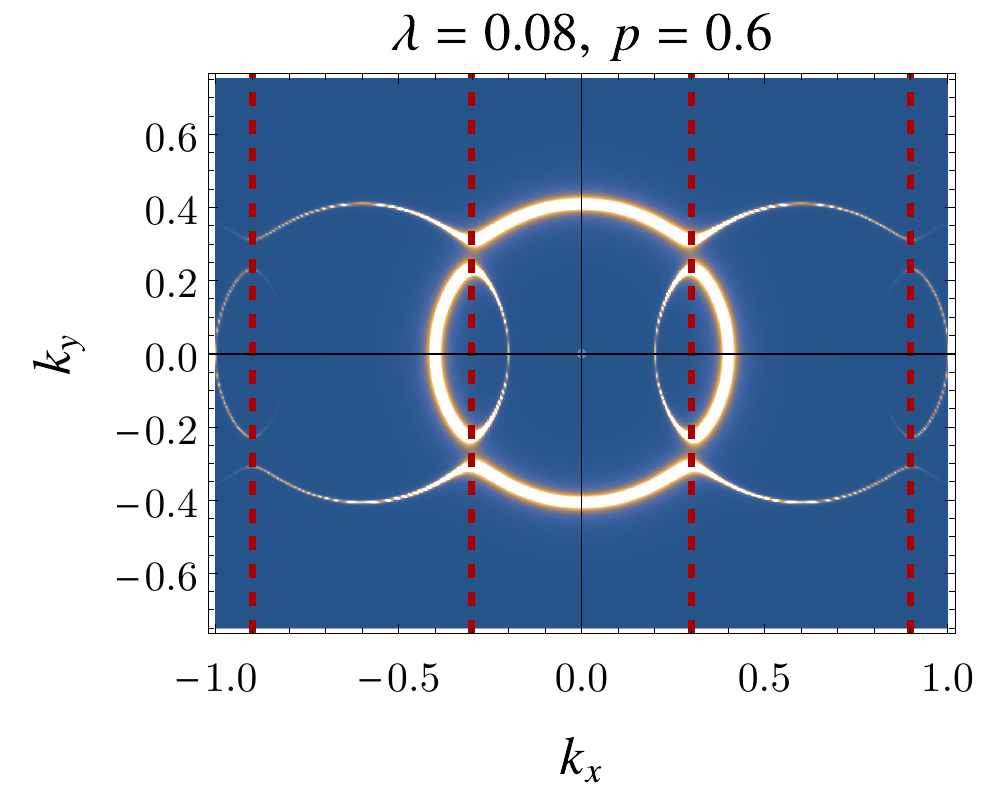} & \includegraphics[width=\myWidR\linewidth]{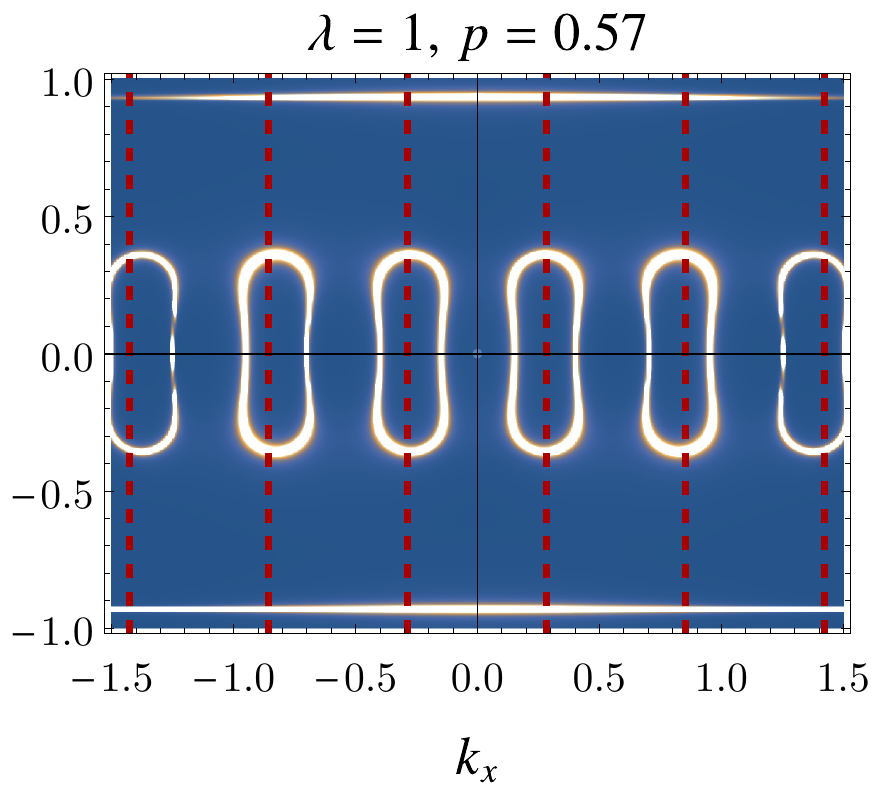}
\end{tabular}
\caption{\label{fig:toy_umklapp} \small{\textbf{The shape of the Fermi surface of a non-interacting 2+1 dimensional Dirac toy model in a unidirectional periodic chemical potential.}  Shown is the spectral density \eqref{equ:spect_density_lattice}, which equals the 00 component of the inverse matrix $M$ \eqref{equ:vector_toy_dirac}. \\
\textit{Left column:} At weak modulation amplitude $\lambda=0.08$, the shape of the Fermi surface appears circular as in the absence of a lattice. However, lattice copies appear in the neighboring Brillouin zones (BZ). When the BZ (red dashed gridlines) becomes smaller than the Fermi momentum, these copies overlap and an Umklapp gap is opened at the intersection point, giving rise to Fermi pockets. \\
\textit{Right column:} At strong modulation amplitude $\lambda=1$ nonlinear effects are seen. The shape of the Fermi surface is now affected even far from the boundary of the BZ. When BZ gets smaller, the FS is first squeezed and then strong Umklapp gaps are opened, leading to the small dumbbell-like Fermi pockets near $k_x$ axis  and the nearly $k_x$ independent flat ``band'' at finite $k_y$.}}
\end{figure}

In Fig.\,\ref{fig:toy_umklapp}, we show various examples of the spectral density in the toy model \eqref{eq:toy_Dirac} for various parameters of the background potential.\footnote{The Green's function of the real equation \eqref{equ:vector_toy_dirac} is real. In order to make the spectral function visible, we evaluate it a slightly imaginary frequency $\omega = E_f + i 10^{-4}$, $E_f =0.4$.}
In the left column we consider a weak periodic potential ($\lambda=0.08$) for three values of the size of the Brillouin zone $p$. For a weak periodic potential all the results can be easily understood from linear analysis. When the BZ is large enough, the Fermi surface does not reach the BZ boundary and has the same isotropic shape as in the case without modulation with Fermi momentum $k_f$. For a smaller BZ (smaller $p$, middle panel), lattice copies of the Fermi surface from the neighbouring Brillouin zones are seen. These copies are visible in the $G_{00}$ component due to the off-diagonal mixing terms $i \lambda \mu/2$ in Eq. \eqref{equ:vector_toy_dirac}. Therefore, the spectral density in these lattice copies is suppressed by a factor of $\lambda$ and, in case of the weak potential, they are hardly visible. When the Brillouin zone size is further reduced ($p/2<k_f$), the neighbouring Fermi surfaces start to overlap and Umklapp gaps are opened at the Brillouin zone boundary. From the point of view of the inversion of matrix \eqref{equ:vector_toy_dirac}, this is simply a linear eigenvalue repulsion effect due to non-zero off-diagonal terms $i \lambda \mu/2$. This occurs around the point when the eigenvalues of the top-left and bottom-right blocks of the matrix \eqref{equ:vector_toy_dirac} become identical. In this case, the series of circular Fermi surfaces turn into a series of Fermi pockets in addition to a an outer band parallel to $x$-axis. This ``band'' is an artefact of the unidirectional modulation of the potential, which we introduced for simplicity. In the case of a realistic crystal lattice in both $x$- and $y$- directions, Fermi pockets would form in the $y$- direction as well.

For strong modulating potential $\lambda = 1$, right column of Fig.\,\ref{fig:toy_umklapp}, the situation is more involved. Even though qualitatively the response is the same, quantitatively non-linear effects will now also affect the shape of the Fermi surface. Firstly, of course, the lattice copies of the Fermi surface acquire a larger spectral weight due to the stronger mixing and are visible already for large Brillouin zones (top row). As the BZ decreases, the strong interaction effects that change the shape of the Fermi surface are seen. It becomes ``squeezed'' as in the middle row of  Fig.\,\ref{fig:toy_umklapp}. Finally, once the Fermi surfaces overlap, the umklapp gap that opens is so large that the outer ``band'' gets pushed far away and is almost flat (independent of $k_x$), while the heavily deformed Fermi pockets are stretched along the BZ boundary, deforming in a dumbbell-like shape, see the bottom row of Fig.\,\ref{fig:toy_umklapp}.  

As we will see below, these different types of the Fermi surface geometries that we find in the toy model \eqref{eq:toy_Dirac} with non-interacting electrons on top of the periodic potential will also appear in the fermionic response of the strongly coupled holographic model. This will aid us in distinguishing effects that are due to strong self-interactions from effects that are due to strong lattice potential.

\section{\label{sec:holographicFS}Holographic Fermi surfaces and zeros}

Next we recall the universal presence of zeros in the fermionic spectral response in a finite-density holographic model of a strongly interacting system of fermions. This is also so in the absence of a lattice and the simplest example of such a system -- the homogeneous model at finite temperature and chemical potential -- is described by a Reissner-Norstr\"om black hole in the 3+1 dimensional curved AdS space \cite{Zaanen:2015oix}. The corresponding metric reads
\begin{equation}
\label{equ:RN_metric}
\mbox{RN black hole} \qquad ds^2 = \frac{1}{z^2}\left[- f(z) dt^2 + dx^2 + dy^2 + \frac{dz^2}{f(z)} \right]
\end{equation}
with
\begin{equation}
f(z) = \left(1 - \frac{z}{z_h} \right) \left( 1  +  \frac{z}{z_h} + \frac{z^2}{z_h^2}  -  \frac{\mu^2 z_h^2}{4} \frac{z^3}{z_h^3} \right) \equiv \left(1 - \frac{z}{z_h} \right) P\left(\frac{z}{z_h}\right)
\end{equation}
and $z_h$ being the radius of the black hole horizon which is related to the temperature and chemical potential in the dual theory:
\begin{equation}
\label{equ:temperature}
16 \pi T z_h = 12 - \mu^2 z_h^2.
\end{equation}
The gauge field potential of the charged black hole is simply
\begin{equation}
A_t = (z-z_h) \mu.
\end{equation}
In what follows we will set $z_h=1$ by choosing the appropriate measuring units.

In order to study the fermionic response in the holographic framework, one considers the Dirac equation on the curved space-time \eqref{equ:RN_metric}  \cite{Liu:2009dm,Iqbal:2011ae,Faulkner:2011tm,Faulkner:2009wj,Iqbal:2009fd,Cubrovic:2009ye,Cubrovic:2011xm,Zaanen:2015oix}:
\begin{equation}
\label{equ:Dirac_RN}
\left[\Gamma^f \bm{e}_f^\mu \left( \p_\mu + \frac{1}{4} \bm{\omega}^a_{b\mu} \eta_{ac} \sigma^{cb} - i q A_\mu \right) - m \right]\Psi(t,x,y,z) = 0,
\end{equation}
where $\Psi = (\psi^\ua, \psi^\da)^T$ is a 4-component 3+1 dimensional Dirac spinor, which we break into two  2-component subparts corresponding to the different spin states on the boundary.
Here $\bm{e}^f_\mu$ are the tetrad vectors; $\bm{\omega}^a_{b\mu}$ is the spin connection\footnote{Spin connection is defined as via $\partial_{\mu}\bm{e}^a_{\nu}+\bm{\omega}^a_{b\mu}\bm{e}^b_{\nu} -\bm{\Gamma}^{\tau}_{~\mu\nu}\bm{e}_{\tau}^a=0$, where $\bm{\Gamma}^{\tau}_{~\mu\nu}$ is Christoffel symbol.}; 
$\eta_{ac}$ is the Minkowski $(-,+,+,+)$ metric; $A_\mu$ is the gauge field, and
$\sigma^{ab}\equiv \frac{1}{2} [\Gamma^a,\Gamma^b]$ is the generator of Lorentz transformations, where we choose the $\Gamma$-matrices to be
\begin{equation}
\label{equ:gamma_matr_herzog}
\Gamma^{\underline{t}} = i \mathbb{1} \otimes \sigma^1,\quad \Gamma^{\underline{x}} = \sigma^3 \otimes \sigma^2 , \quad \Gamma^{\underline{y}} = \sigma^2 \otimes \sigma^2,\quad \Gamma^{\underline{z}} = \mathbb{1} \otimes \sigma_3.
\end{equation}
The tunable parameters here are $q$ and $m$ -- the charge and mass of the bulk fermion field. 
In the homogeneous background \eqref{equ:RN_metric} one can immediately expand in plane waves along $\{x,y\}$ directions. It is also convenient to rescale the spinor and introduce the new fields $\zeta^a$:
\begin{equation}
\label{eq:spinor_rescaling}
\psi^a = (g/{g_{zz}})^{1/4} e^{i k_x x + i k_y y - i \omega t} \zeta^a(z)\,, \qquad a\in \{\uparrow, \downarrow\}
\end{equation}
where $g_{\mu \nu}$ is the metric and $g$ its determinant.
After this redefinition, the equation for $\zeta^{\ua}$ reads
\begin{equation}
\label{equ:Dirac_RN_expanded}
\left[\p_z - \frac{m}{z \sqrt{f}} \sigma_3 + i \frac{\omega + q A}{f} \sigma_2 + \frac{k_x}{\sqrt{f}} \sigma_1\right] \zeta^{\ua}(z) - i k_y \frac{1}{\sqrt{f}}\sigma_1  \zeta^{\da}(z)= 0.
\end{equation}
Taking advantage of the isotropy of  the RN black hole, we  choose the coordinates in such a way that $k_y = 0$ and the Weyl spinors  $\zeta^\ua$ and $\zeta^\da$ decouple. For $\zeta^{\da}$, the $k_x$ momentum term has the opposite sign. Therefore, in the subsector $k_y=0$ the two spinors describe left- and right-moving modes in the $x$-direction. This simplification generically doesn't happen in the presence of a lattice, however it still arises when the fermion propagates along the unidirectional potential, as we will see in the next section\footnote{This is a consequence of our choice of the $\Gamma$-matrices \eqref{equ:gamma_matr_herzog}. Their reduction on the boundary, defined by the action of 2+1 dimensional Lorentz generator $\sigma^{\mu \nu}$, $\mu,\nu \in (t,x,y)$ on a positive subspace of $\Gamma^z$ coincides with $\bar{\gamma}^\mu$ in \eqref{eq:toy_Dirac}. The different spin states which we consider are the eigenstates of $\bar{\gamma}^y$ and they decouple in case when the fermions propagate along $x$-axis, since in the corresponding 1+1 dimensional problem they possess different chiralities.}.

The analysis near the AdS boundary $z\rar 0$ shows  the spinor components behave as 
\begin{equation}
\label{eq:boundary_expansion}
\zeta^{\ua}(z;\omega, k)\big|_{z \rar 0} = 
\begin{pmatrix}
b^{\ua}(\omega,k) z^m + a^{\ua}(\omega,k) \mathfrak{a}^0_r (\omega,k) z^{-m+1} + \dots\\ a^{\ua}(\omega,k) z^{-m} + b^{\ua}(\omega,k) \mathfrak{b}^0_s (\omega,k) z^{m+1} + \dots
\end{pmatrix},
\end{equation}
where the constants $\mathfrak{a}^0_r, \mathfrak{b}^0_s$ can be fixed by solving the equations of motion, see Appendix~\ref{app:RN_fermions}, and can in principle depend on parameters $(\omega,k)$.
Given the expansion \eqref{eq:boundary_expansion} of the bulk fermion profile one identifies the independent coefficients $a$ and $b$ with respectively the source and the expectation value of the corresponding fermionic boundary operator $\Psi$ of conformal  dimension $\Delta = \frac{3}{2} + m$ \cite{Contino:2004vy,Cubrovic:2009ye,Laia:2011wf, Iqbal:2011ae,Liu:2009dm,Faulkner:2009wj,Faulkner:2010tq}. This is called ``direct quantization'' (Direct Q). It is important to note however, that this identification is not unique: for  $m \in [0,\tfrac{1}{2})$ one can consider an ``alternative quantization'' (Alt.~Q, which we denote with tilde: $\tilde{\Psi}$) and consider $b$ as the source and $a$ as the expectation value of the operator $\tilde{\Psi}$ with conformal dimension $\tilde{\Delta} = \frac{3}{2} - m$. In this regime of $m \in [0,\tfrac{1}{2})$, a single bulk model can correspond to the two distinct boundary theories, depending on the type of quantization chosen. A useful way of studying alternative quantization is to consider the direct framework but to extend the range of $m$ to negative values of $m \in (-\tfrac{1}{2},0]$. In this way the roles  of $a$ and $b$ are  switched in \eqref{eq:boundary_expansion}. 

Within the linear response approximation, the holographic identification of source and response allows one to obtain the two-point function of the fermionic operator under consideration: the fermionic Green's function $G = \la \Psi^{\dagger} \Psi \ra= (\mathrm{response})/(\mathrm{source})$. In general the Green's function is a matrix, and to divide out the sources requires a few steps. In the notation of \cite{Iqbal:2009fd}, given the relation between the sources and responses\footnote{Note that in the boundary fermionic theory the $a^\ua$ is a source to $b^\da$ and vice versa, see \cite{Benini:2010pr,Benini:2010qc}}
\begin{equation}
\label{equ:S_matrix}
\begin{pmatrix}
b^{\ua} \\ b^{\da}
\end{pmatrix}
=
\mathcal{S} \begin{pmatrix}
a^{\da} \\ a^{\ua}
\end{pmatrix},
\end{equation}
the (retarded) fermionic Green's function 
reads 
\begin{equation}
\label{equ:G_from_S}
G_{R} = - i \mathcal{S} \gamma^{\underline{t}},
\end{equation}
where $\gamma^{\underline{t}} = i \sigma_1$ with  our choice of the $\Gamma$-matrices \eqref{equ:gamma_matr_herzog} \cite{Iqbal:2009fd,Cubrovic:2009ye,Liu:2009dm, Faulkner:2009wj,Benini:2010pr,Benini:2010qc}.
Therefore, in direct quantization  $G^{\ua \ua} = b^{\ua}/a^{\ua}$, while in alternative quantization  $\tilde{G}^{\ua \ua} = a^{\ua}/b^{\ua}$. In other words, the two point functions in both quantizations are related by
\begin{equation}
\label{equ:RN_Green}
 G^{\ua \ua} = b^{\ua}/a^{\ua} = 1/\tilde{G}^{\ua \ua}. 
\end{equation}
In particular and importantly, this means that the poles of the Green's function in the alternatively-quantized boundary theory correspond to the zeros of the Green's function in case of the direct quantization.  

The final point is to determine which type of the Green's function we are looking at. This is fixed by the boundary conditions at the black hole horizon, see Appendix \ref{app:RN_fermions}. The retarded Green's function corresponds to a purely infalling solution at the horizon: 
\begin{equation}
\label{eq:infalling}
\zeta^\ua(z) \Big|_{z\rar z_h} \sim (z_h - z)^{- i \omega / 4 \pi T}. 
\end{equation}

In short, in order to evaluate the Green's function, one has to solve the Dirac equation \eqref{equ:Dirac_RN_expanded} in the full bulk geometry \eqref{equ:RN_metric} and find the ratio between the $a$ and $b$ branches of the AdS boundary expansion \eqref{eq:boundary_expansion} of the solution.
The first-order ordinary differential equation \eqref{equ:Dirac_RN_expanded} together with the boundary condition \eqref{eq:infalling} is solved with the numerical shooting method, as we explain in Appendix \ref{app:RN_fermions}, and for given values of the background parameters $\{T,\mu\}$, fermionic parameters $\{q, m\}$, frequency and momentum $\{\omega,k_x=k\}$, the asymptotic form \eqref{eq:boundary_expansion} is read off from the solution. This gives us the Green's function $G^{\ua \ua}(\omega, k)$ from Eq. \eqref{equ:RN_Green}. In complete analogy we can evaluate $G^{\da \da}$ by solving the appropriate equation, or simply use the symmetry mentioned above   $G^{\da \da}(k) = G^{\ua \ua}(-k)$. Finally, we  evaluate the fermionic spectral density as
\begin{equation}
\label{equ:spectral density}
 A(\omega, k) = \mathrm{Im} \, \mathrm{Tr} G (\omega, k) = \text{Im}( G^{\ua\ua}(\omega, k)+G^{\da\da}(\omega, k))\,,
 \end{equation}
which is the central object of our study. 
In what follows we mostly focus on the features of the Fermi surface, defined as the locus of the spectral density peaks at zero frequency: $\omega(k_F)=0$. 

\begin{figure}
\center
\includegraphics[width=0.49 \linewidth]{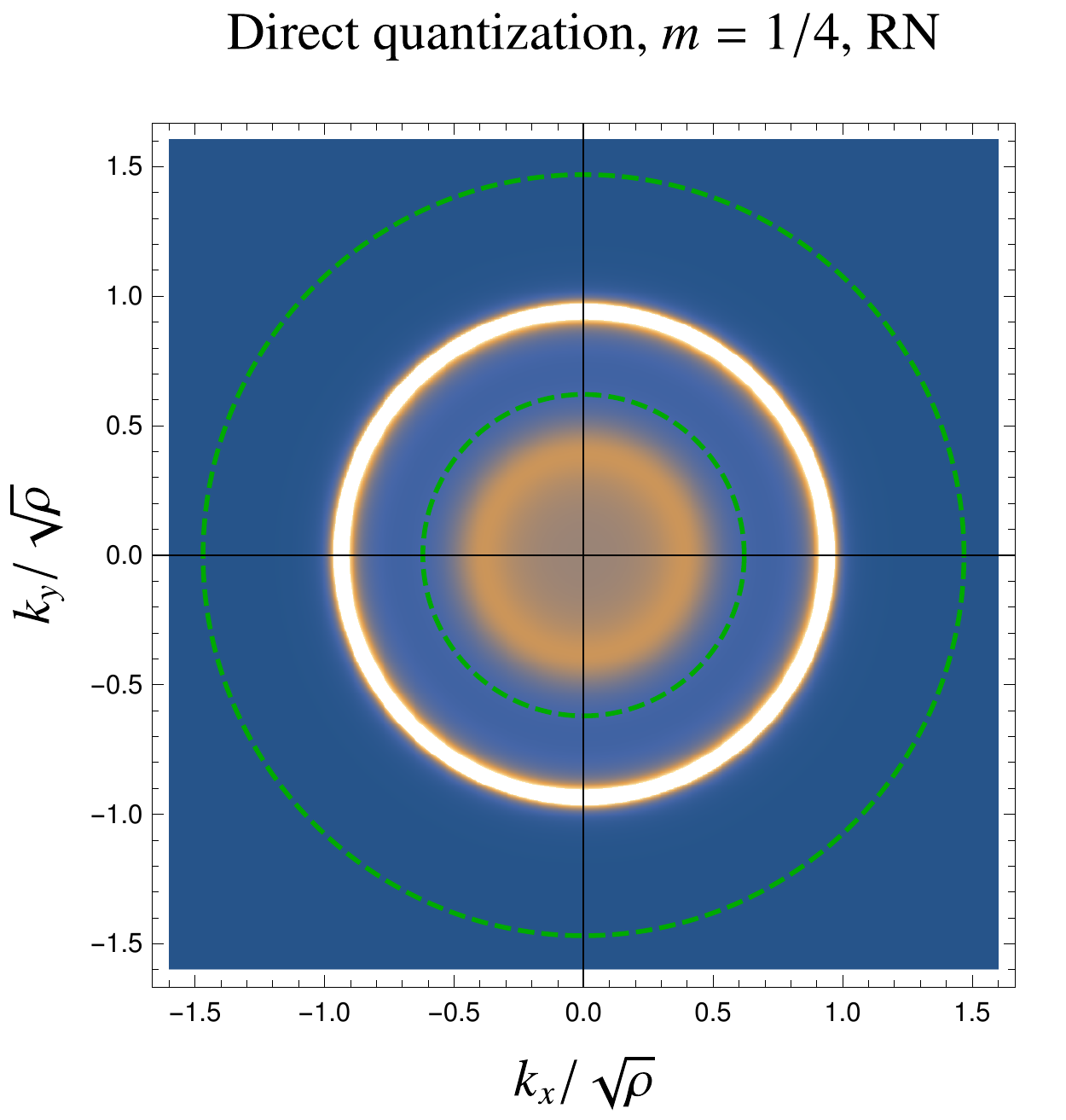}
\includegraphics[width=0.49 \linewidth]{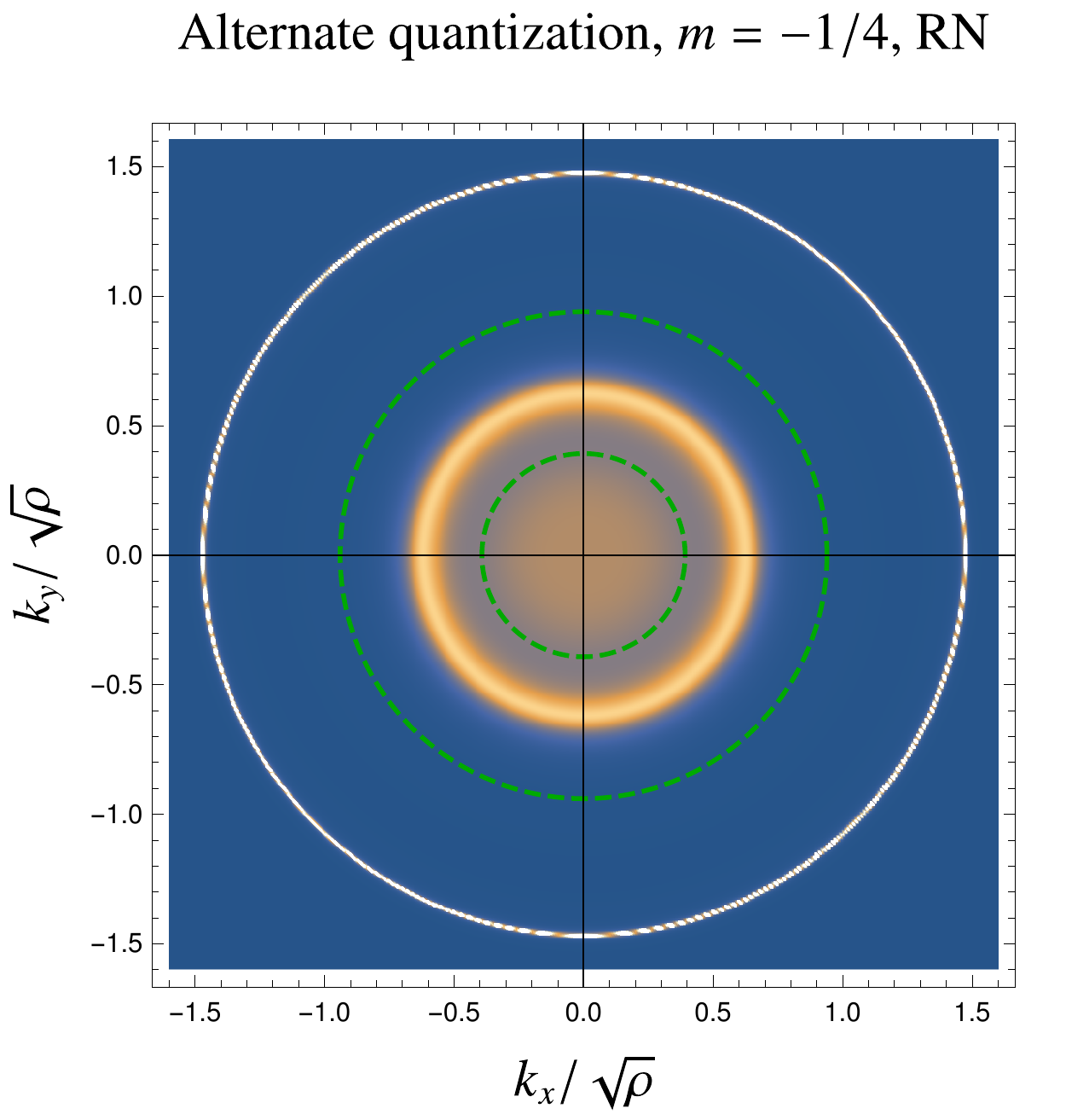}\\
\caption{\label{fig:RN_fermions_density} \textbf{The fermionic spectral density at near the Fermi level ($\omega=0.01 \sqrt{\rho}$) in the isotropic Reissner Nordstr\"om background \eqref{equ:RN_metric} for $m=1/4, \, q=1$}. The density plot shows the distribution of the spectral density in momentum plane $k_x$, $k_y$ for the direct (left panel) and alternative (right panel) quantization picture. The circular Fermi surfaces are seen, which are expected for the isotropic background. In both cases there are two nested FS with different lifetime of the excitation. The positions of the FS in the other quantization picture are shown with the dashed lines on each panel, which makes apparent their alternating structure. $T/ \sqrt{\rho} =0.01$, $\mu^2/\rho \approx 3.329$}
\end{figure}

The fermionic spectral density \eqref{equ:spectral density} at zero frequency in the RN black hole \eqref{equ:RN_metric} computed this way is shown on Fig.\,\ref{fig:RN_fermions_density}.
We have chosen the fermionic bulk mass $m=1/4$, the charge $q=1$ and $T/\mu \approx 0.005$. The Fermi surface with direct quantization is shown on the left panel. Moreover, since $m<1/2$ alternative quantization is also possible and the result is shown on the right. The Fermi surfaces are circular, as expected for fermionic excitations at finite chemical potential in an isotropic background. We also observe the specific holographic feature of the appearance of multiple nested Fermi surfaces \cite{Faulkner:2009wj}. This is a generic  feature in holographic models: the number of the observed Fermi surfaces depends on the mass and the charge of the bulk fermion as well as on the background chemical potential, and in principle it can be arbitrary \cite{Zaanen:2015oix}. 

\begin{figure}[t!]
\center
\includegraphics[width=0.49 \linewidth]{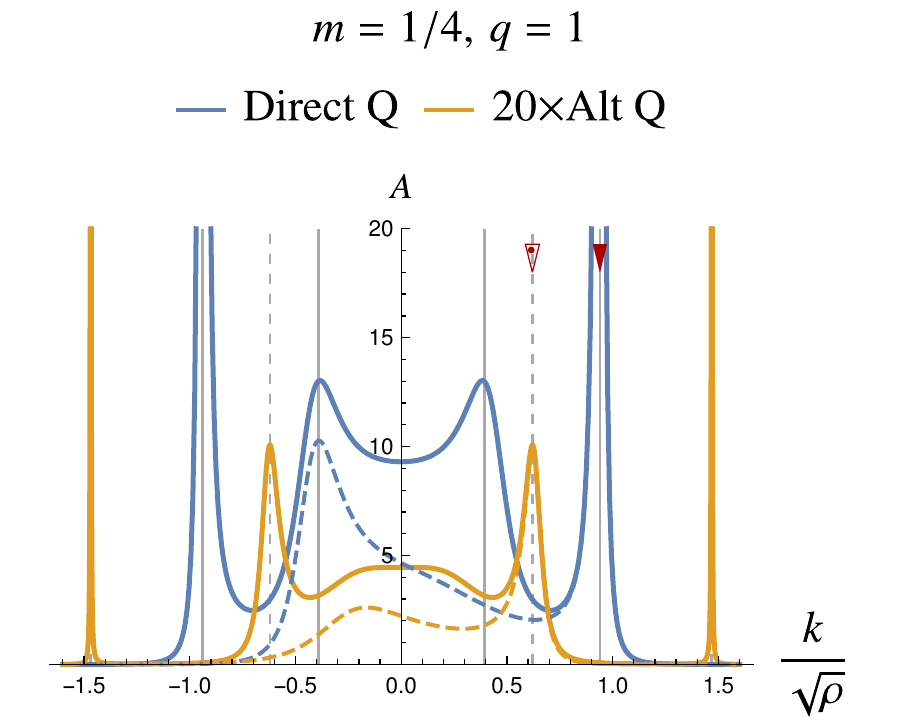}
\includegraphics[width=0.49 \linewidth]{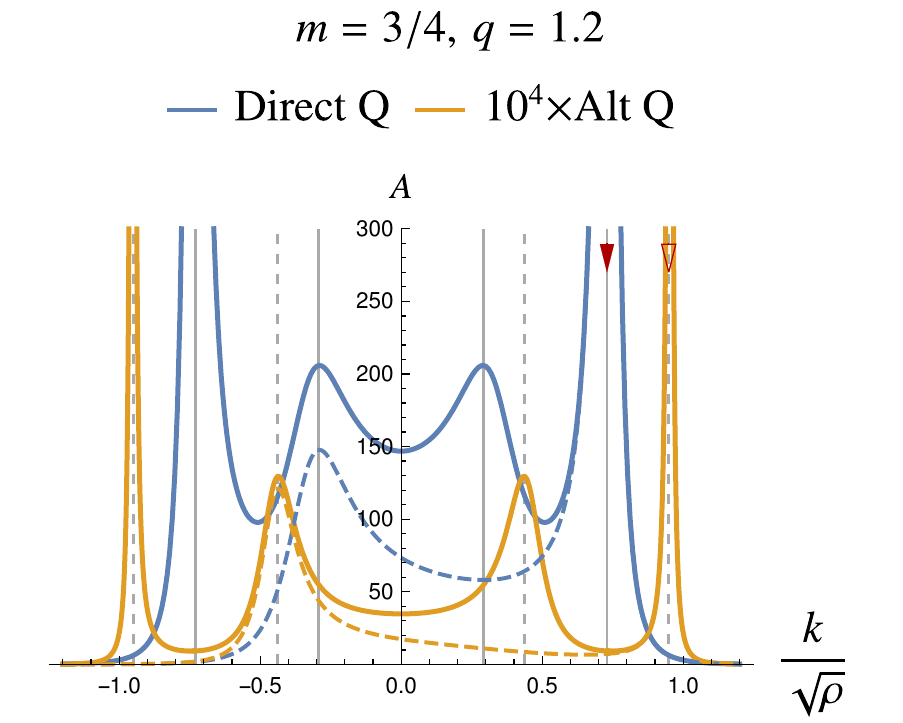}\\
\caption{\label{fig:RN_fermions} \textbf{The fermionic spectral density near the Fermi level ($\omega=0.01 \sqrt{\rho}$) in the isotropic Reissner-Nordstr\"om background \eqref{equ:RN_metric}.} The $k_y=0$ cuts in linear scale of the direct and alternative quantization pictures are shown. The left panel corresponds to the fermionic mass $m=1/4 < 1/2$, where the alternatively quantized dual CFT is well defined, while the right panel corresponds to the case $m=3/4 > 1/2$, where it does not exist. In both cases the clear peaks corresponding to the well defined Fermi surfaces are seen in both direct and alternative pictures (In case $m=3/4$ one needs larger value of fermionic charge $q=1.2$ in order for the FS to be formed). The dashed lines show the contribution of a single spin component to the spectral density. The red arrows point out the peaks which, if brought close to each other, may be destroyed due to the proximity of the zero and a pole in the given spin Green's function. In all cases $T/ \sqrt{\rho} =0.01$, which matches the parameters we will use later on.
}
\end{figure}

Here we wish to call attention to
another interesting phenomenon which is evident when  comparing the results of  direct and alternative quantizations: each plot in Fig.\,\ref{fig:RN_fermions_density} includes the position of the Fermi surface in the other quantization in dashed lines. We see that the Fermi surfaces in the direct and alternative quantization alternate. This effect is more clearly visible  on the left panel of Fig.\,\ref{fig:RN_fermions}, where the $k_y=0$ cut of the spectral density is shown. Indeed, the solid and dashed vertical grid lines, indicating the Fermi surfaces in the direct and alternative quantization, respectively, alternate: there is always an alternative quantization FS in between two direct ones.

At this point it is important to remember  that, due to the  inverse relation between the Green's function in Direct Q. and Alt.\,Q. \eqref{equ:RN_Green}, a pole in the alternative Green's function $\tilde{G}$ always corresponds to a zero in the Direct Green's function $G$. While a pole near the real axis produces a discernible peak in the spectral density $A$, a zero is only reflected in a depletion of spectral density with a minimum set by the imaginary part of the position of zero in the complex plane, which is proportional to temperature. This depletion, unlike the peaks, is harder to spot in density plots like Fig.\ref{fig:RN_fermions_density}. 
The reason is simply that in order for the depletion in spectral density to be visible, the latter must have a finite background value. However, the spectral density at zero frequency is set by temperature itself and therefore the depletion is not seen. Therefore, the most direct way of detecting the zero in the Green's function is indeed to study the peaks in its alternatively quantized counterpart. 
This is how we will identify the zeroes of the Green's function in the remainder of this paper.

Another reason for the absence of detectable depletions of the spectral density is that the trace of the Green's function contains several additive terms, only one of which is suppressed. Indeed, in a particularly simple example with $k_y=0$ the Green's function is diagonal in the spin representation
\begin{equation}
\mathrm{ Im} \, \mathrm{Tr} G (\omega, k) = \mathrm{ Im}  G^{\ua \ua} + \mathrm{ Im}  G^{\da \da} = \mathrm{ Im}  \left[\frac{1}{\tilde{G}^{\ua \ua}} \right] + \mathrm{ Im}  \left[\frac{1}{\tilde{G}^{\da \da}} \right]\,.
\end{equation}
Therefore, the peaks of a single component, shown in dashed in Fig.\,\ref{fig:RN_fermions}, are clearly seen in the total $\mathrm{ Im} \, \mathrm{Tr} G$, while the depletion, if any, would be shaded with the finite value coming from the opposite-spin component. One could alternatively capture the position of zeros by looking at the real part of the Green's function at real $\omega$, which changes sign at exactly this point. However this method suffers from the same problem: the contribution of the two spin components add together in the trace of the Green's function and one has to diagonalize it in the spin space in order to distill the position of zeros.

The Green's function zeros become particularly important  in the  case where the peaks in both direct and alternative quantization of the same spin component (shown with arrows in Fig.\,\ref{fig:RN_fermions}) come closer to each other. In this situation, the pole and the zero of the Green's function in the complex $\omega$ plane would recombine and the residue of the pole would vanish 
In this way, a peak in the spectral density would be ``eaten'' by the approaching depletion producing a distinct observable phenomenon. This pattern is similar to what happens in a Fano resonance in a continuum coupled to a discrete system, but here it is manifested as a destruction of the Fermi surface at $\omega=0$. In the simple isotropic model described above this does not happen since the poles and zeros are always separated. In what follows we will show that the situation changes when a strong periodic background chemical potential is considered.

Before moving forward, 
another comment regarding the existence of the zeros in the Green's function is in order. On the right panel of Fig.\,\ref{fig:RN_fermions} we show the results obtained for a different mass $m=3/4$ for which only a single quantization in the dual CFT is possible \cite{Cubrovic:2009ye,Laia:2011wf}. 
In other words, one cannot prescribe a physical meaning to the Alt.\,Q Green's function. Formally, however, we can still evaluate the Alt.\,Q expression by inverting the Direct Q result. The artificially computed Alt.\,Q spectral density clearly exhibits the peaks even in this case, which therefore correspond to the zeros of the Direct Q Green's function. Therefore, we conclude that the general mechanism of appearance of the zeros, is independent of whether or not the alternatively quantized picture is well defined.

\section{\label{sec:model}Fermionic spectral function in a holographic lattice}

We now turn to the study on how the fermionic spectral function discussed in Sec.\,\ref{sec:holographicFS} is affected by the periodic chemical potential, and in particular when the strength of the lattice potential is strong.
The bulk model with periodically modulated chemical potential -- the holographic ionic lattice -- has been introduced in \cite{Horowitz:2012gs} and studied in great detail in \cite{Donos:2014yya,Rangamani:2015hka}. In what follows we will adhere to the framework used previously by some of us in \cite{Krikun:2017cyw,Mott,Andrade:2017leb}. We consider the holographic model with Einstein-Maxwell action 
\begin{equation}
\label{equ:Einstein_Maxwell_action}
	S = \int d^4 x \sqrt{- g} \left( R  - \frac{1}{4} F^2 - 2 \Lambda \right),\qquad \Lambda=-3,
\end{equation}
where $F=dA$ is the gauge filed strength tensor, $R$ is the Ricci scalar and $\Lambda$ -- the cosmological constant. We introduce the spatially modulated chemical potential (c.f. \eqref{eq:mux})
\begin{equation}\label{eq:mux_hologrpahy}
	A_t(x,z)\Big|_{z=0} = \mu(x) = \mu_0 [1 + \lambda \sin(p x) ]\,.
\end{equation}
The non-isotropic black hole solution can now no longer be constructed analytically, but must be found numerically.
In order to find the gravitational background solution it is sufficient to consider the following metric ansatz
\begin{align}\label{ds2 anstaz}{}
	ds^2 &= \! \frac{1}{z^2}\left( \! - \mathcal{T}^2 f(z) dt^2 + \mathcal{Z}^2 \frac{dz^2}{f(z)} + \mathcal{X}^2 (dx + Q_{zx} dz)^2 + \mathcal{Y}^2 dy^2  \right), \\
\notag
	{\cal A} &= A_t dt, 
\end{align}
The blackening factor $f(z)$ is that of the RN black hole \eqref{equ:RN_metric}
and all the other ansatz functions  depend  on both $z$ and $x$ coordinates. Given that, at the horizon \hbox{$(\mathcal{Z}^2-\mathcal{T}^2)\big|_{z=1}= 0$}, the temperature is still given by \eqref{equ:temperature}. Using the DeTurck trick \cite{Wiseman:2011by,Adam:2011dn,Headrick:2009pv} and the  numerical methods for solving partial differential equations (PDEs) developed in \cite{Mott,Krikun:2017cyw,Krikun:2018ufr} we obtain the background gravitational solutions for a given temperature $T$ at a fixed chemical potential. The lattice wave-vector is  $p$ and $\lambda$ is the amplitude of the chemical potential modulation in units of $\mu_0$. The details of the numerical procedure and precision control are discussed in Appendix\,\ref{app:numerics}. With the gravitational background solution at hand we proceed to solve the Dirac equation \eqref{equ:Dirac_RN}. The co-frame is now less trivial than for the isotropic case:
\begin{align}
\label{equ:tetrad}
\bm{e}_{\underline{t}} = \frac{\sqrt{f} \mathcal{T}}{z}  dt, && \bm{e}_{\underline{x}} = \frac{\mathcal{X}}{z} (dx + Q_{zx} dz), &&
\bm{e}_{\underline{y}} = \frac{\mathcal{Y}}{z}  dy, && \bm{e}_{\underline{z}} = \frac{\mathcal{Z}}{\sqrt{f} z} dz,
\end{align} 
but we can still follow the procedure outlined in Sec.\,\ref{sec:holographicFS}. Since we are now working in a periodic background potential, we express the spinors in  Bloch waves \eqref{eq:toy_Dirac} instead of plane waves and therefore the rescaling of Eq. \eqref{eq:spinor_rescaling} is modified as
\begin{equation}
\label{equ:holographic_bloch}
\psi^a(z, x) = (g/{g_{zz}})^{1/4} e^{i k_x x + i k_y y - i \omega t} \zeta^a(z, x), \quad  \zeta^a(z, x + 2 \pi/p) \equiv \zeta^a(z, x), \quad a\in \!\{\uparrow, \downarrow\},
\end{equation}
where $\zeta^a(z, x)$ is now a position dependent periodic Bloch wave function.
Substituting \eqref{equ:tetrad} and \eqref{equ:holographic_bloch} into the Dirac equation \eqref{equ:Dirac_RN}, we obtain the fermionic equations of motion, which are now PDEs with $(2 \pi/p)$ - periodic boundary conditions in $x$-direction (c.f. \eqref{equ:Dirac_RN_expanded}):
\begin{gather}
\begin{split}
\label{equ:Dirac_lattice}
& \bigg[ \p_z  - \frac{m}{z \sqrt{f}} \mathcal{Z} \sigma_3  + i \frac{\omega + q A}{f} \frac{\mathcal{Z}}{\mathcal{T}} \sigma_2 + \frac{k_x}{\sqrt{f}} \frac{\mathcal{Z}}{\mathcal{X}} \sigma_1\bigg]  \zeta^{\ua}(x,z) - i k_y \frac{1}{\sqrt{f}}\frac{\mathcal{Z}}{\mathcal{Y}} \sigma_1  \zeta^{\da}(x,z)
 \\
 & - \left[ i  \frac{1}{\sqrt{f}}\frac{\mathcal{Z}}{\mathcal{X}} \left(\p_x + \frac{1}{4}\p_x \ln \frac{1}{\sqrt{f}}\frac{\mathcal{Z}}{\mathcal{X}}  \right) \sigma_1 
+ Q_{zx} \left(\p_x + \frac{1}{2}\p_x \ln Q_{zx} + i k_x \right) \bm{1} \right] \zeta^{\ua}(x,z)
= 0.
\end{split} 
\end{gather}
The equation for the $\zeta^{\da}$ component is obtained by a parity transformation: $k_i \rar -k_i$, $\p_i \rar - \p_i$ and $Q_{zi} \rar -Q_{zi}$, for $i \in \{x,y\}$.
The AdS boundary and horizon asymptotic behavior \eqref{eq:boundary_expansion}, \eqref{equ:boundary_series}, \eqref{eq:infalling}, \eqref{eq:horizon_series} remain unchanged in the presence of the periodic potential, except that all the expansion coefficients are now periodic functions of the boundary coordinate~$x$. The details of the numerical algorithm used to integrate these equations are given in Appendix \ref{app:fermi_numerics}. 

Given a solution of \eqref{equ:Dirac_lattice} as a function of both $z$ and $x$,  we extract the near boundary coefficients $a(x)$ and $b(x)$ defined in Eq.\,\eqref{eq:boundary_expansion}. In order to obtain the linear response $\mathcal{S}$-matrix of Eq.\,\eqref{equ:S_matrix}, we expand these periodic expansion coefficients in Fourier series similarly to the method shown in Sec.\,\ref{sec:umklapp}:
\begin{equation}
\label{equ:fourier_reps}
a(x) = \sum a_l e^{i l p x}, \qquad b(x) =  \sum b_n e^{i n p x}.
\end{equation}
Therefore, the $\mathcal{S}$-matrix is a infinite matrix with both spin and Fourier indices (c.f. \eqref{equ:vector_toy_dirac}).
\begin{equation}
\label{equ:lattice_S_matrix}
\begin{pmatrix}
b^{\ua}_n \\ b^{\da}_n
\end{pmatrix}
=
\sum_{l} \mathcal{S}_{n l} \begin{pmatrix}
a_l^{\da} \\ a_l^{\ua}
\end{pmatrix},
\end{equation}
The Green's function is evaluated in the same way as in Eq.\,\eqref{equ:G_from_S}, except that it is now a matrix. 
It is worth mentioning that, since the Fourier basis exponents differ from each other by a shift with exactly one unit of the lattice wave vector $p$, the indices $m,l$ can be interpreted as the Brillouin zone index. Therefore, the Green's function $G_{ml}$ is simply a matrix in the Brillouin zone representation. We comment on some features of its structure in Appendix\,\ref{app:BZ}. 

As it was discussed earlier in Sec.\ref{sec:umklapp}, the most interesting component of the Green's function for us is $G_{00}$, which measures the overlap between the response function in the material and the plane waves of the ARPES probe \eqref{equ:spect_density_lattice}.  In order to measure $G_{00}$ and the associated spectral density, we consider a plane wave source $a(x) = 1$ as a boundary condition when solving the bulk equations of motion \eqref{equ:Dirac_lattice} and read off the homogeneous component of the response $b(x)$, see also Appendix\,\ref{app:fermi_numerics}.

In order to evaluate the Green's function in the alternative quantization picture, one has to invert the full infinite matrix and the simple formula \eqref{equ:RN_Green} turns into
\begin{equation}
\label{equ:lattice_alt_G}
\tilde{G}_{00} = \left(G^{-1}\right)_{00}.
\end{equation}
In order to perform this inversion one has to evaluate all the components of the Direct~Q Green's function, which is hard in practice even when one truncates the Fourier series. Therefore, instead of using this method we evaluate the alternatively quantized Green's function by solving the Dirac equation with the negative fermionic mass parameter (see Appendix\,\ref{app:RN_fermions}). As explained above, this switches the roles of $a(x)$ and $b(x)$ in the linear response calculation. In Appendix\,\ref{app:BZ} we check that the two approaches give identical results.

\section{\label{sec:result}Destruction of the Fermi surface by the zeros in the Green's funciton}
Let us now analyze the results which we get for the fermionic spectral function in the holographic model described in Sec.\,\ref{sec:model}. In what follows we study a series of the background gravitational solutions for various Brillouin zone sizes $p$, but with fixed charge density $\rho$. We also fix the temperature $T/\sqrt{\rho},$\footnote{In practice we have control over the mean chemical potential $\mu_0$, which is set by the boundary condition, but we can tune it in such a way that the mean charge density, which can be read out for a given background solution stay fixed when we change the lattice wave-vector $p$.} and the amplitude of the chemical potential modulation~$\lambda$ \eqref{eq:mux_hologrpahy}. We  consider two cases with weak and strong potential modulation, in direct analogy with the toy model study performed in Sec.\,\ref{sec:umklapp}. We use the same temperature and  parameters for the bulk fermion as in the homogeneous RN-black hole case addressed in Sec.\,\ref{sec:holographicFS}:
\begin{equation}
\frac{T}{\sqrt{\rho}} = 0.01, \qquad m = 1/4, \qquad q =1.
\end{equation}
As seen in Fig.\,\ref{fig:RN_fermions_density}, the size of the Fermi surface in the absence of the periodic potential for these parameters is $k_f \approx 0.94 \sqrt{\rho}$.

\begin{figure}[t!]
\center
\begin{tabular}{c}
\includegraphics[height=7cm]{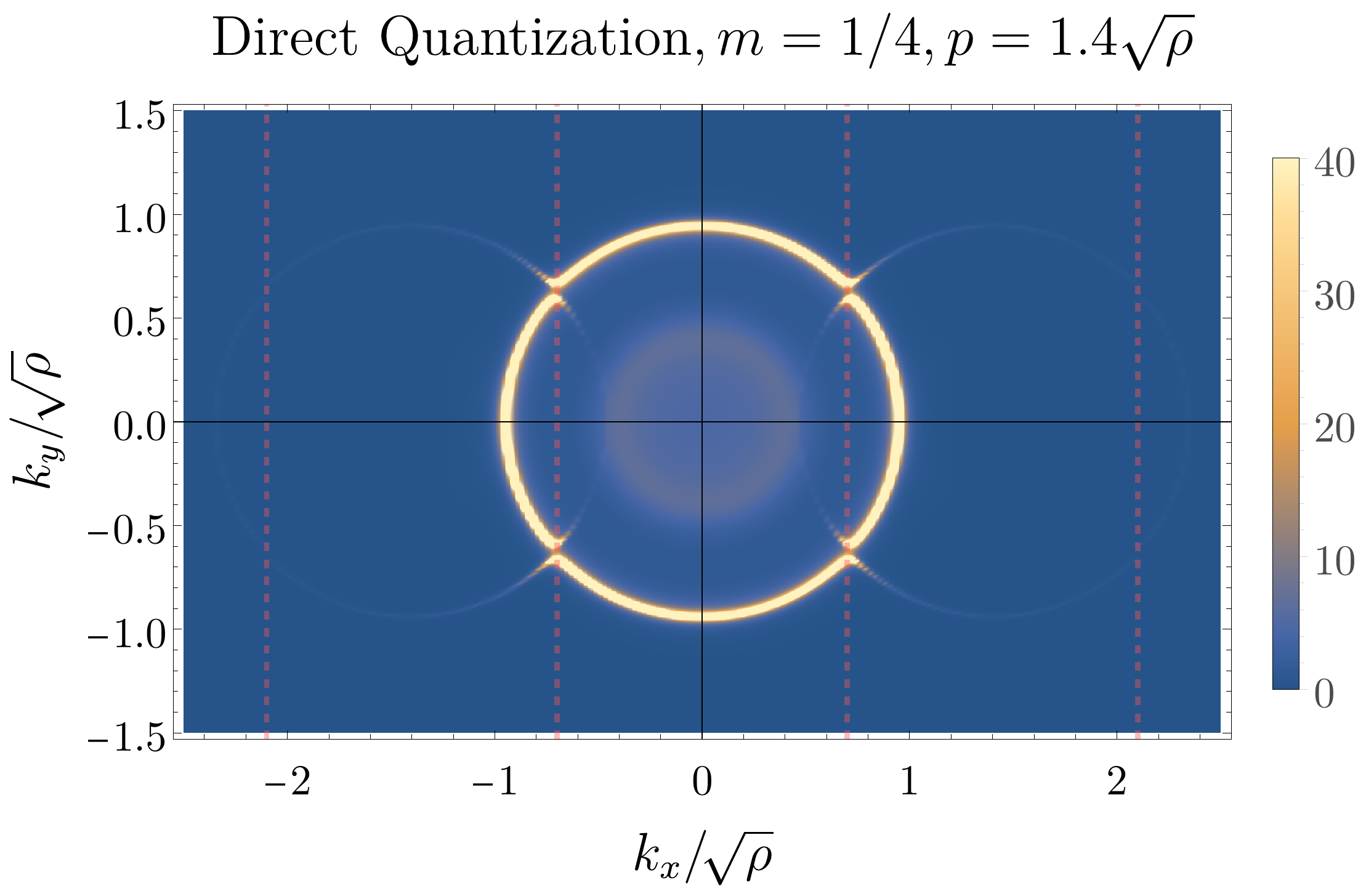} 
\\
\includegraphics[height=7cm]{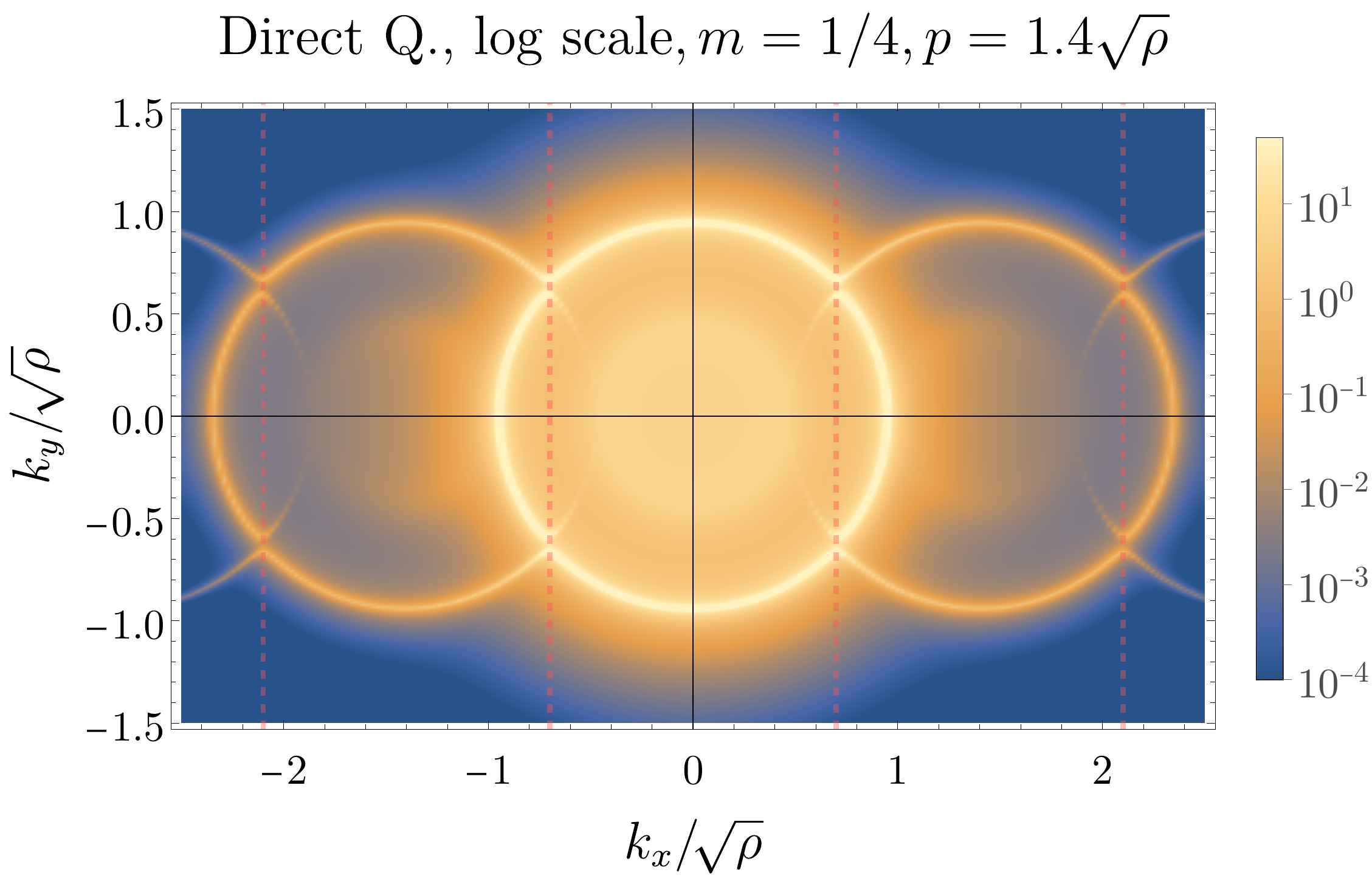}
\end{tabular}
\caption{\label{fig:lattice_FS_weak_coupling} \textbf{The fermionic spectral density in the holographic model with weak unidirectional periodic potential ($\lambda = 0.1$).} \\
The density plots at small ($\omega = 0.01\sqrt{\rho}$) show the effects of introducing the periodic potential. The two plots represent the same data on linear (upper) and logarithmic (lower) scales. Near the Brillouin zone boundary, indicated by the dashed lines, the umklapp gap is starting to open. The weak nature of the potential suppresses the intensity of the copies in neighbouring Brillouin zones, to the extent that they are hardly visible in the upper plot. The lower plot shows that the copies are clearly visible on a logarithmic scale. 
\\
The background parameters are: $p=1.4\sqrt{\rho}$, $\lambda = 0.1$, $T = 0.01\sqrt{\rho}$, $\mu^2 \approx \rho/0.3$.
}
\end{figure} 

\begin{figure}[t!]
\center
\begin{tabular}{c}
\includegraphics[width=0.8 \linewidth]{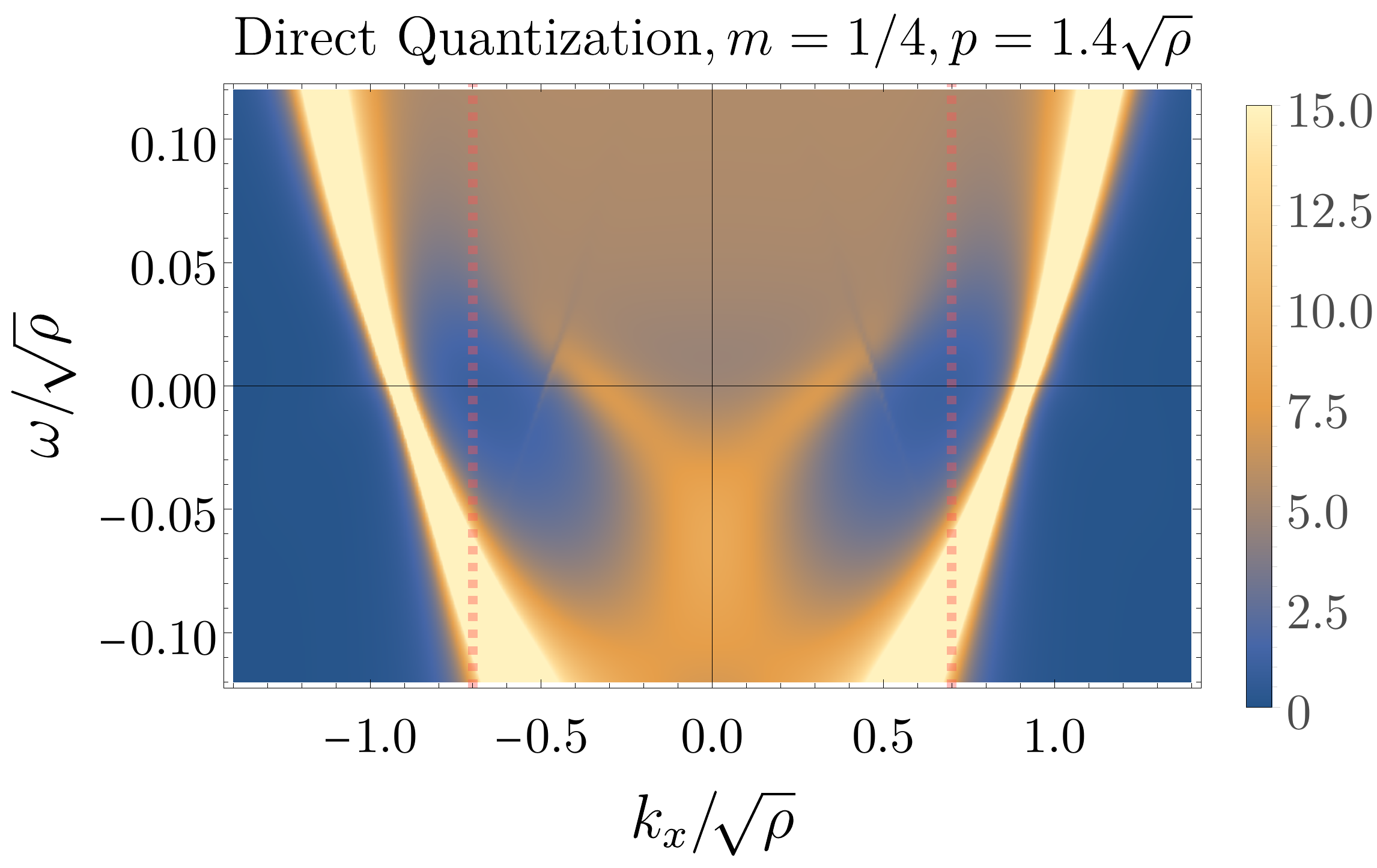}
\end{tabular}
\caption{\label{fig:lattice_FS_weak_coupling_EDC} \textbf{The frequency resolved fermionic spectral density in the holographic model with weak unidirectional periodic potential ($\lambda = 0.1$).} \\
The density plot shows a cut through the same data as Fig.\ref{fig:lattice_FS_weak_coupling} in the $k_x,\omega$ plane, at $k_y = 0.001 \sqrt{\rho}$. Around $(k_x,\omega) \approx (\pm 0.4, 0.02)$, the dark lines crossing through the inner Dirac cone arise from the "resonant" type zeroes \cite{Gnezdilov:2018qdu} discussed in Sec.\,\ref{sec:Intro}.
}
\end{figure} 

\subsection{Weak lattice potential}

We start with  a weakly modulating lattice with strength $\lambda=0.1$. 
On Fig.\,\ref{fig:lattice_FS_weak_coupling}, we show the momentum distribution of the spectral density, which in this case displays the circular Fermi surface with exactly the same size as the homogeneous one (c.f. Fig.\,\ref{fig:RN_fermions_density}). 
The weak lattice potential does not affect the shape of the Fermi surface. 
We choose the size of the Brillouin zone to be smaller then the Fermi momentum $p < 2 k_f$, therefore the umklapp copies of the FS overlap and, in perfect agreement with the observations in toy-model of Sec.\,\ref{sec:umklapp}, we see  umklapp gaps opening at the BZ boundaries.
Another feature which is expected is the suppression of the spectral density in the neighboring Fermi surfaces. 
Indeed, these are almost invisible in the linear scale plot on top of Fig.\,\ref{fig:lattice_FS_weak_coupling}. However, the bottom panel shows the logarithm of the spectral density, which makes the $\lambda$-suppressed lattice copies of the Fermi surface clearly visible. 
In a nut shell, the results obtained in the holographic model with a weak periodic potential are in perfect agreement with both the standard logic of fermion physics in a periodic potential, and the shape and size of the holographic Fermi surface in the homogeneous background. 
This serves as a consistency check of our approach and numerical techniques.     
   
Similarly, a familiar pattern is also seen on the energy-momentum resolved spectral function shown on Fig.\,\ref{fig:lattice_FS_weak_coupling_EDC}. 
As expected, the usual holographic dispersion relation for  fermionic excitations is recovered. In this case, it consists of the two nested Dirac cones in the vicinity of the Fermi level. 
The spectral lines quickly broaden due to the quantum critical self energy $\Sigma(\omega) \sim \omega^{2\nu_{k_F}}$ \cite{Faulkner:2009wj,Cubrovic:2009ye,Faulkner:2011tm,Zaanen:2015oix}. 
However, there is one distinctive new feature due to the presence of the holographic lattice. There is a localized depletion of the spectral weight along the lines which correspond to the dispersion bands of the neighbouring Brillouin zones.
This feature is a consequence of a general effect pointed out earlier in \cite{Gnezdilov:2018qdu}. 
Namely, when multiple systems with discrete spectra interact with each other by means of the quantum critical continuum, the spectral function of one of the  systems develops isolated zeros, or depletions, at the positions of the energy levels of the other systems. 
In the case of the holographic model with a lattice, the discrete systems are the umklapp copies of the fermionic dispersion while the quantum critical continuum originates from the near horizon geometry \cite{Faulkner:2011tm}. 
This interesting effect may, in principle, also lead to a destruction of the Fermi surface at an isolated point, however it will not be relevant in the present study. 
  
\begin{figure}[!ht]
\center
\newcommand\myheight{5.79}
\begin{tabular}{ll}
\includegraphics[height=\myheight cm]{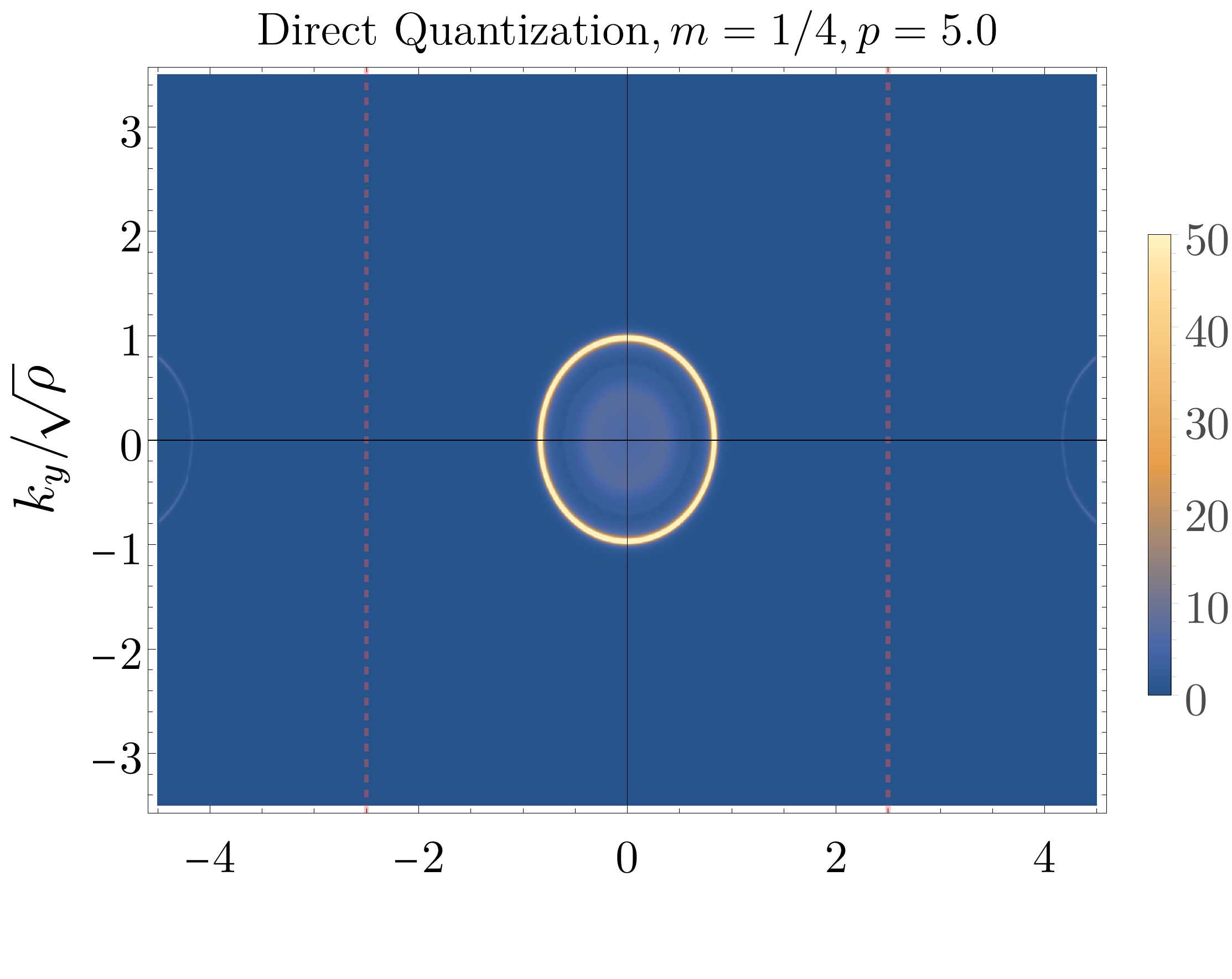} & 
\includegraphics[height=\myheight cm]{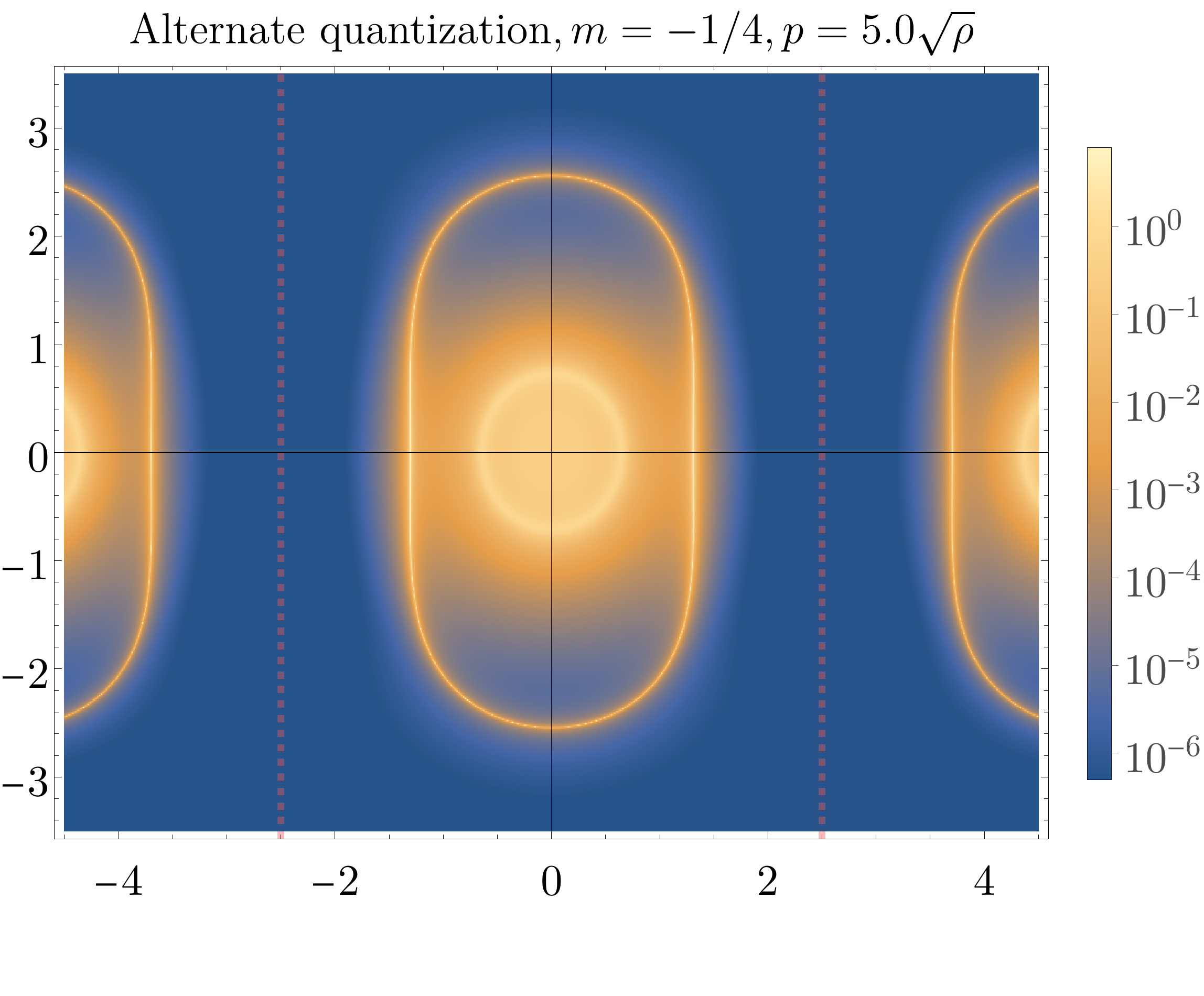}\vspace{-0.5cm}\\
\includegraphics[height=\myheight cm]{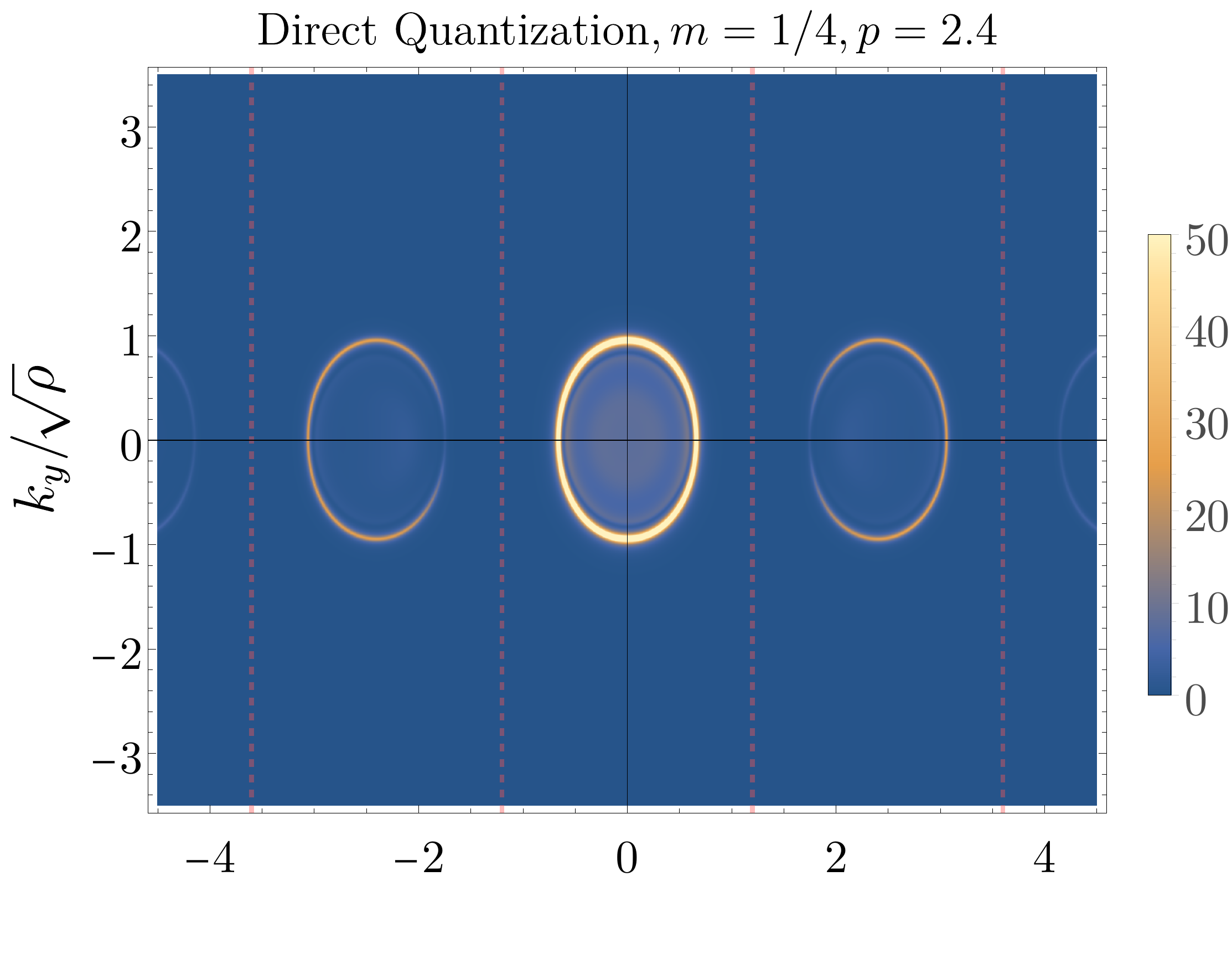} &
\includegraphics[height=\myheight cm]{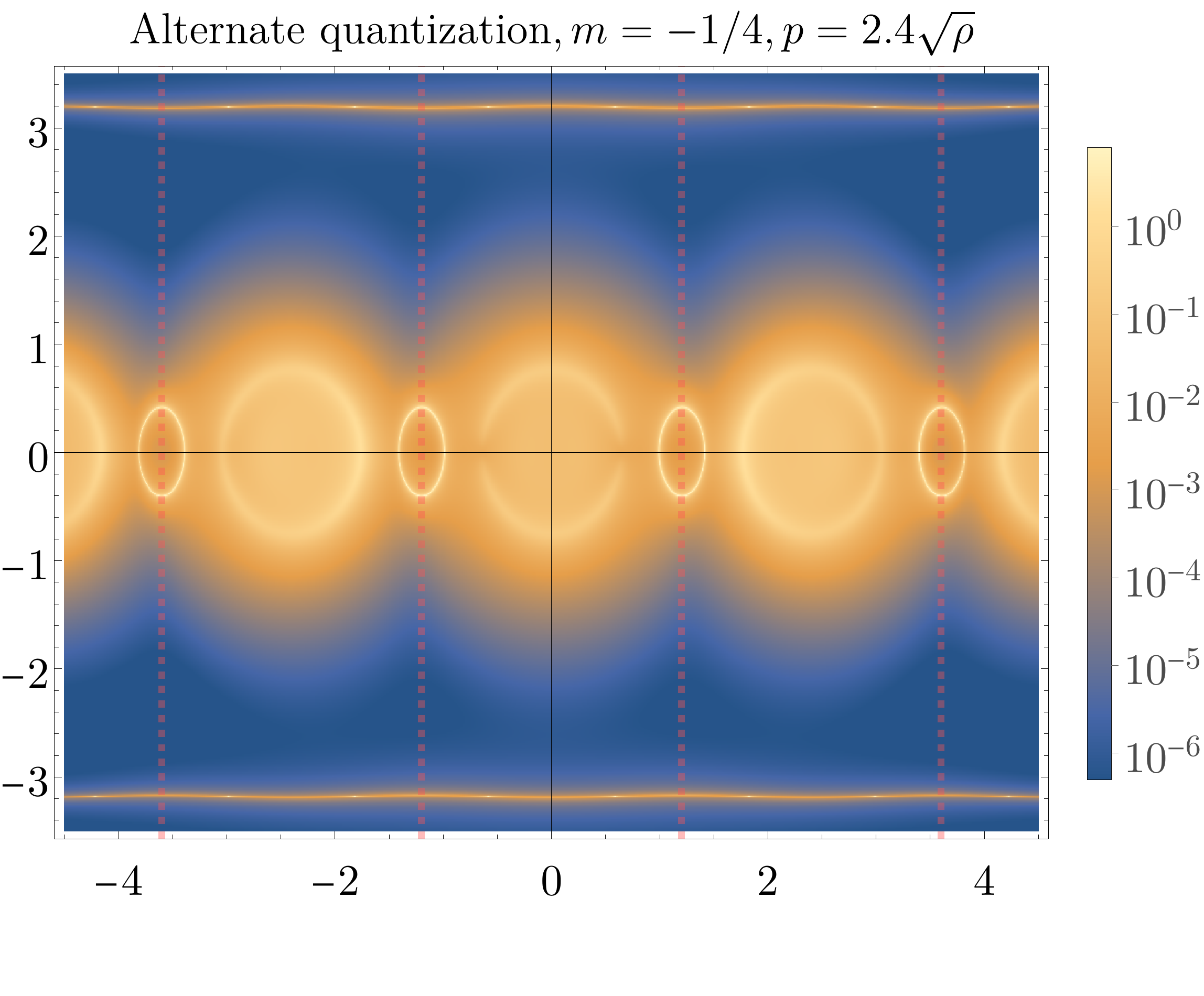}\vspace{-0.5cm}\\
\includegraphics[height=\myheight cm]{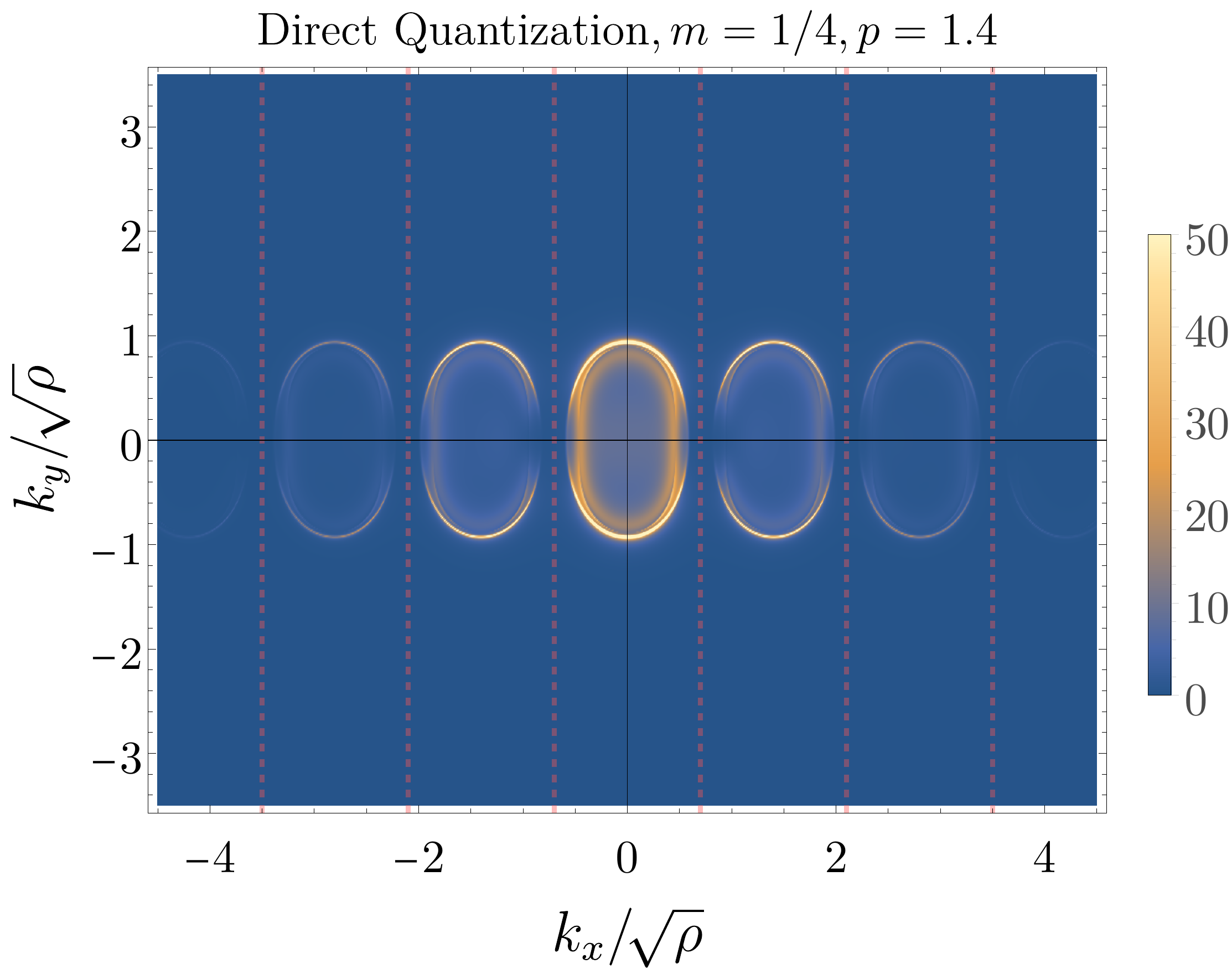} &
\includegraphics[height=\myheight cm]{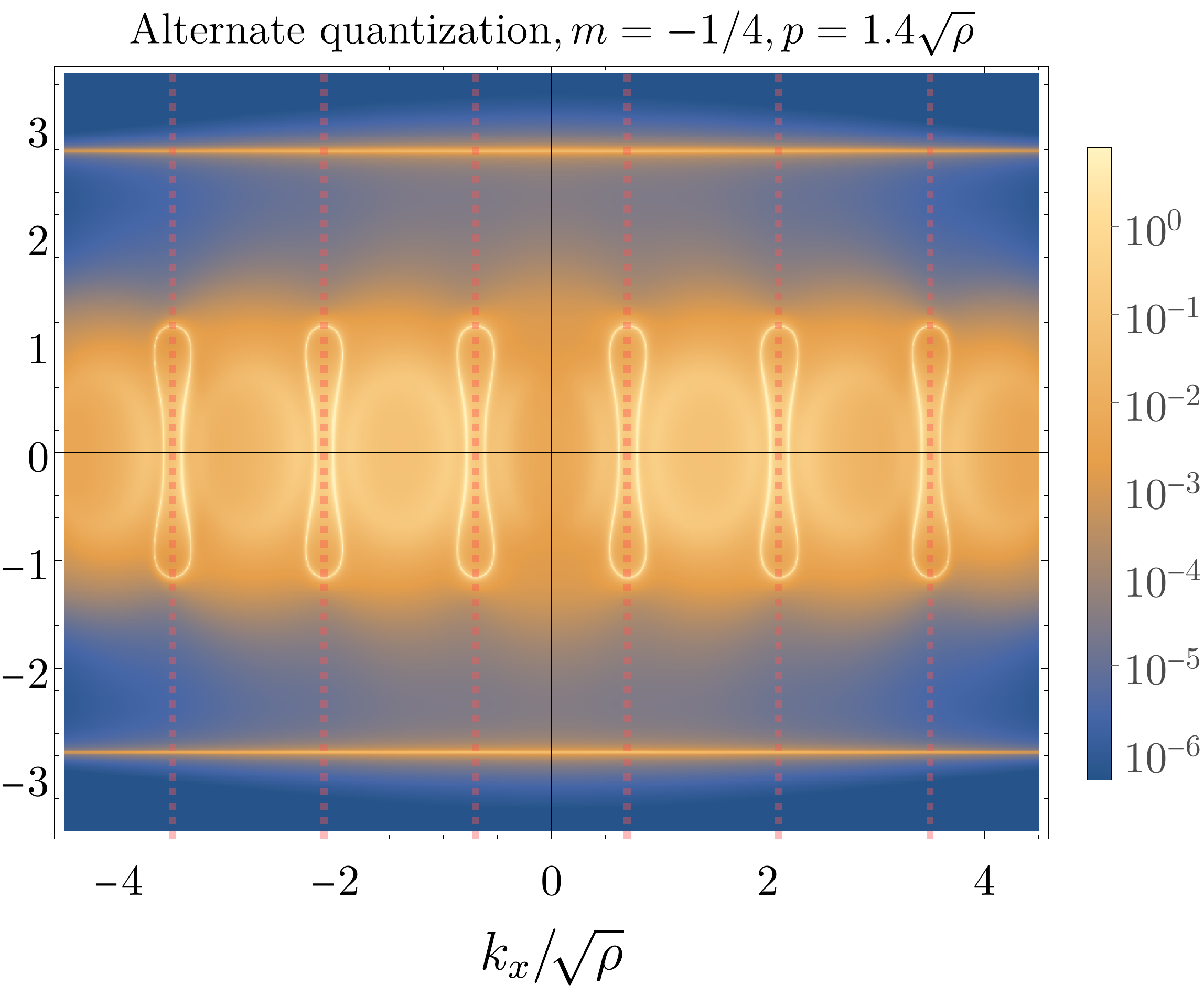}
\end{tabular}
\caption{
\label{fig:lattice_FS_strong_coupling}
\small{
\textbf{The fermionic spectral density in the holographic model with strong unidirectional periodic potential ($\lambda = 8.0$)}, in direct and alternate quantizations near the Fermi level ($\omega=0.01 \sqrt{\rho}$).
The squeezing of the outer Fermi surface happens like in the toy model on Fig.\,~\ref{fig:toy_umklapp}. In the Alt. quantization, where the Fermi surfaces are larger, the outer FS breaks apart in a band at $k_y \approx \pm 3$, and a pocket at the BZ boundary due to Umklapp. When the BZ boundary is brought closer, to $p=1.4\sqrt{\rho}$, the spectral density appears to be drastically reduced in the direct quantization near the Umklapp surface. In the alternate quantization, dumbbell shapes similar to the non-interaction model in Fig.\,\ref{fig:toy_umklapp} appear. 
The background parameters are $T=0.01 \sqrt{\rho}$ and $p=5\sqrt{\rho}, \mu^2 = \rho/0.54$ (top); $p=2.4\sqrt{\rho}, \mu^2 = \rho/1.22$ (middle); $p=1.4\sqrt{\rho}, \mu^2 = \rho/2.04$ (bottom).
}
}
\end{figure}

\subsection{Strong lattice potential}

Let us now turn to the analysis of the case of {strong potential modulation} $\lambda=8$, depicted in Fig.\,\ref{fig:lattice_FS_strong_coupling}. 
Similarly to the toy model study of Sec.\,\ref{sec:umklapp}, we consider a series of backgrounds with different potential wave-vector $p$ which sets the sizes of the Brillouin zone. The parameters considered allow to access the alternative quantization, which has an inverse Green's function $\tilde{G} = G^{-1}$. 
In order to analyze the spectral density in Alt.\,Q, we solve the Dirac equations \eqref{equ:Dirac_lattice} on top of the same gravitational background lattice, but taking the negative sign for the bulk fermion mass $m=-1/4$ (see Appendix \ref{app:BZ} for the details of obtaining the Alt Q fermionic response). We show the results of the direct and alternative quantizations side by side on Fig.\,\ref{fig:lattice_FS_strong_coupling}. Note, however, that we choose logarithmic scale for the Alt.\,Q results in order to resolve all the features of the FS.

On the first row of Fig.\,\ref{fig:lattice_FS_strong_coupling}, the BZ is much larger then the Fermi surface in direct quantization (top left, $p = 5 \sqrt{\rho} \approx 5 k_f$). Therefore, since the umklapp surfaces are far away, the FS is not deformed even in this regime of strong lattice potential.
This is expected; when momentum is much smaller than the BZ, i.e. in the long wave length limit, 
the periodic structure of the potential becomes irrelevant and only the mean value of the chemical potential $\mu_0$ plays a role in the formation of the Fermi surface.\footnote{As is well known by now, this logic can be violated in other holographic models involving homogeneous lattices \cite{Ling:2014bda,Bagrov:2016cnr} or the periodic scalar lattice \cite{bagrovLattice}.} On the other hand, the outer FS in the alternate quantization is larger ($k_f \approx 1.5 \sqrt{\rho}$, see Fig.\,\ref{fig:RN_fermions_density}). Therefore, on the top right panel of Fig.\,\ref{fig:lattice_FS_strong_coupling}, we see that, in this case, the FS is already deformed by the BZ boundaries, similar to the non-interacting toy model in Sec.\,\ref{sec:umklapp}.

When the BZ becomes smaller, $p \approx 3k_f$, second row of Fig.~\ref{fig:lattice_FS_strong_coupling}, the neighbouring Fermi surfaces come close to each other and get deformed due to the strong lattice potential in direct quantization. This is exactly the same situation observed in the toy model of Sec.~\ref{sec:umklapp}. 
 For Alt.\,Q (right panel), we readily observe the formation of  Fermi pockets and the flat outer band. This is again similar to the toy model at strong lattice potential.

\begin{figure}[ht!]
\centering
\includegraphics[width=1\linewidth]{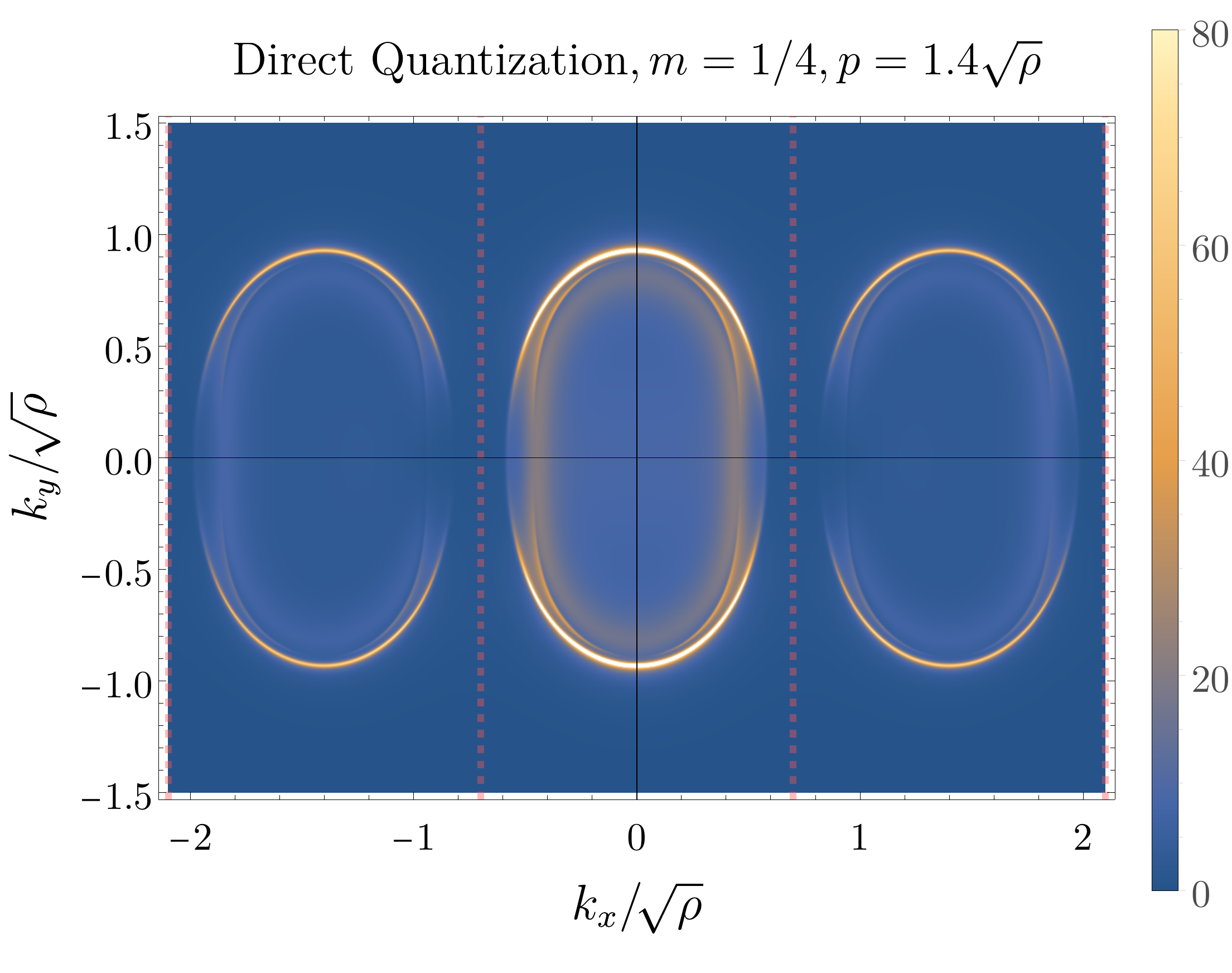}
\caption{\label{fig:lattice_FS_direct_zoom_nolines} \textbf{The destruction of the Fermi surface in strong unidirectional holographic lattice potentials ($\lambda=8.0$)}. Zooming on the picture shown on the bottom left in figure~\ref{fig:lattice_FS_strong_coupling}, it is clearly seen that the Fermi surface loses its sharpness and even disappears around $k_y = 0$.}
\end{figure}

The novel interesting phenomenon arises when the Brillouin zone is squeezed even further. 
On the bottom row of Fig.\,\ref{fig:lattice_FS_strong_coupling} we show the situation with $p=1.4\sqrt{\rho}$. 
Quite strikingly, we see that as the FS is squeezed further more, the sharp spectral density peaks indicating the shape of the Fermi surface disappear along the boundary of the BZ in direct quantization. A more detailed view in Fig.\,\ref{fig:lattice_FS_direct_zoom_nolines} shows this explicitly.
On the other hand, the alternative quantization plot displays heavily squeezed dumbbell-like Fermi pockets, which are centered  at the BZ boundaries.

\begin{figure}[ht]
\begin{tabular}{cc}
\begin{minipage}{0.49\linewidth}
\center
\includegraphics[width=\linewidth]{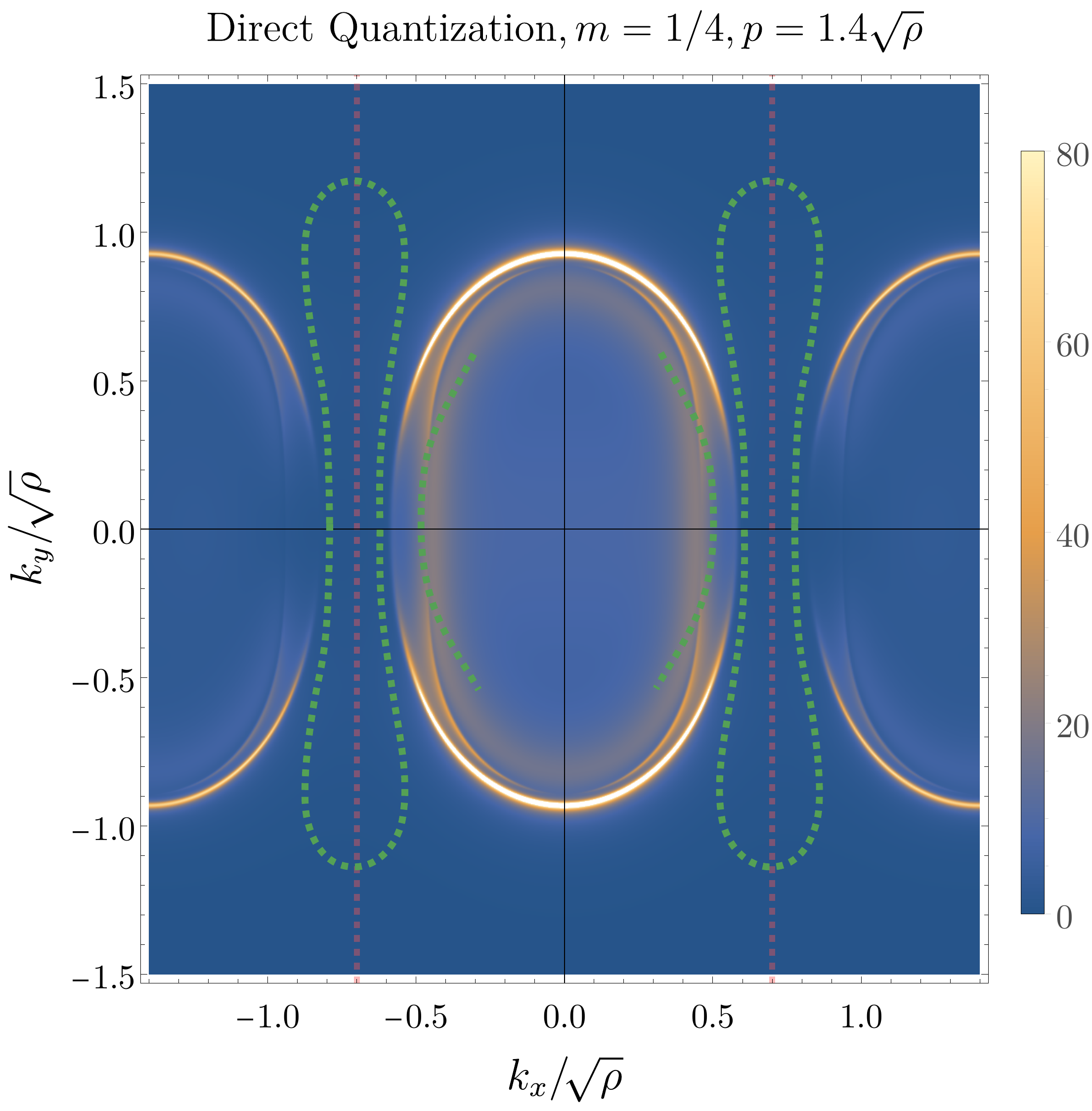}
\end{minipage}
&
\begin{minipage}{0.49\linewidth}
\center
\includegraphics[width=\linewidth]{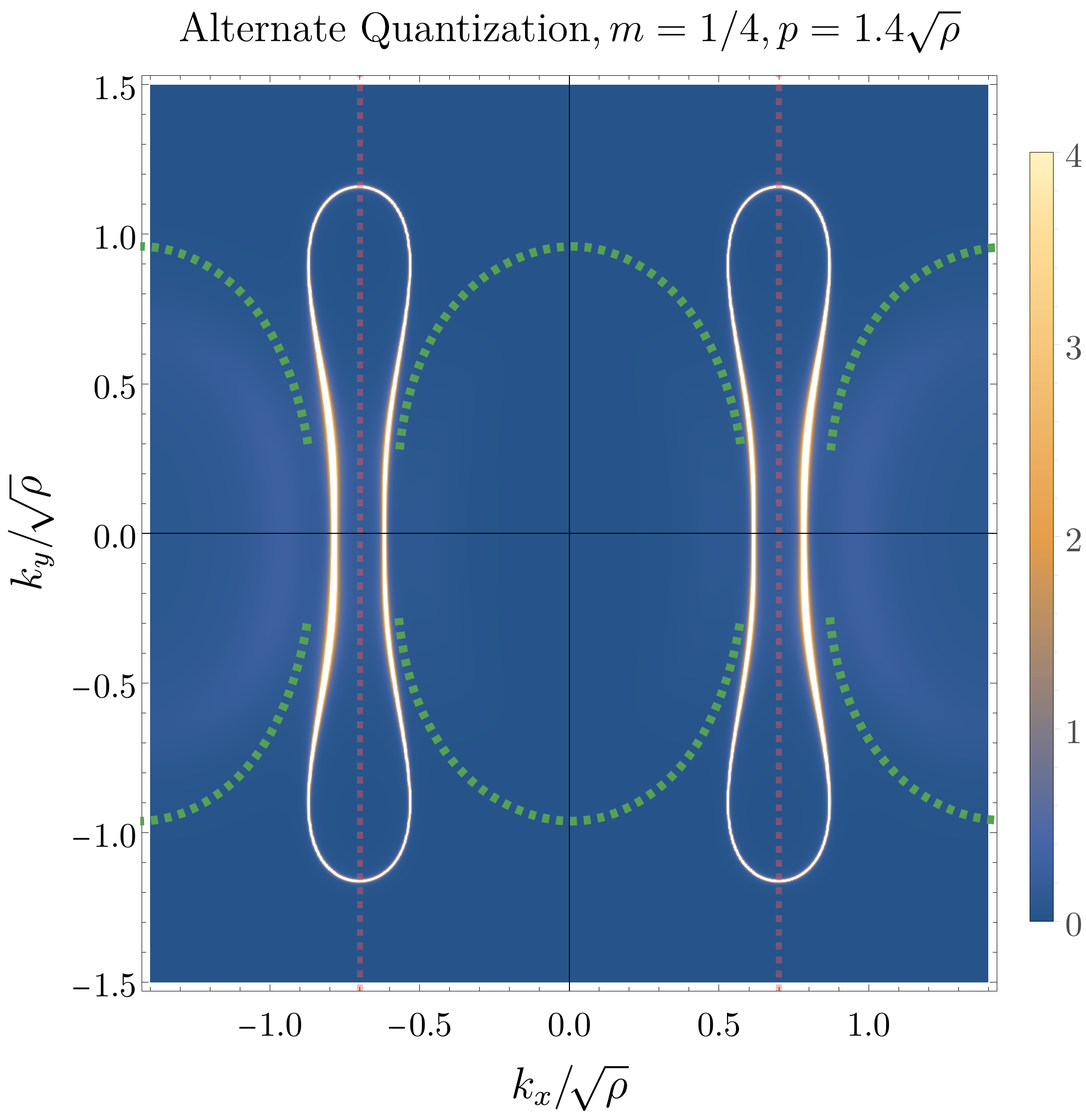}
\end{minipage}
\end{tabular}
\caption{\label{fig:lattice_FS_altQ_compare} \textbf{Poles and zeros of the fermionic response in strong holographic lattices ($\lambda = 8.0$).}
Looking more closely at the spectral densities for large lattices and small BZ ($p=1.4\sqrt{\rho}$), for the direct quantization in the left figure the spectral density broadens and disappears near the Umklapp surfaces $k_x \pm 0.7\sqrt{\rho}$. When tracing out the Fermi surfaces in direct and alternate quantizations and overlaying them on top of each other, it becomes apparent that when the Fermi surfaces of the two different quantization would lie at the same point, they both experience a reduction and broadening of the spectral weight near the FS. The secondary Fermi surface of the alternate quantization is also shown as a trace on the left. This is not visible on the right-hand picture due to the choice of color scheme (see the log-scale Fig.\,\ref{fig:lattice_FS_strong_coupling} instead). 
}
\end{figure}

As discussed in Sec.~\ref{sec:holographicFS}, we have a thorough understanding of some of the peculiar features of the holographic fermionic spectral response. This allows us to figure out the origin of the depletion of spectral function. The  zeros of the direct quantization are the corresponding poles of the alternative quantization Green's function.
In Fig.\,\ref{fig:lattice_FS_altQ_compare}, we show high resolution results of the FS for Direct and Alternative quantizations. In each plot, we show with the dashed green lines the positions of the poles in the other quantization scheme. 
Most strikingly, the position of the secondary Fermi surface in the Alt.\,Q overlaps with the primary Fermi surface in the Direct Q, as seen on the left panel of Fig.\,\ref{fig:lattice_FS_altQ_compare}. Therefore, there are {\em zeros} in Direct Q that get pushed towards the poles defining the FS. This effect destroys the Fermi surface in an extended region near the BZ boundary. This cancellation of poles and zeros is the fundamental reason for the novel phenomenon we observe.

\begin{figure}[t]
\center
\newcommand\myhight{7}
\includegraphics[height=\myhight cm]{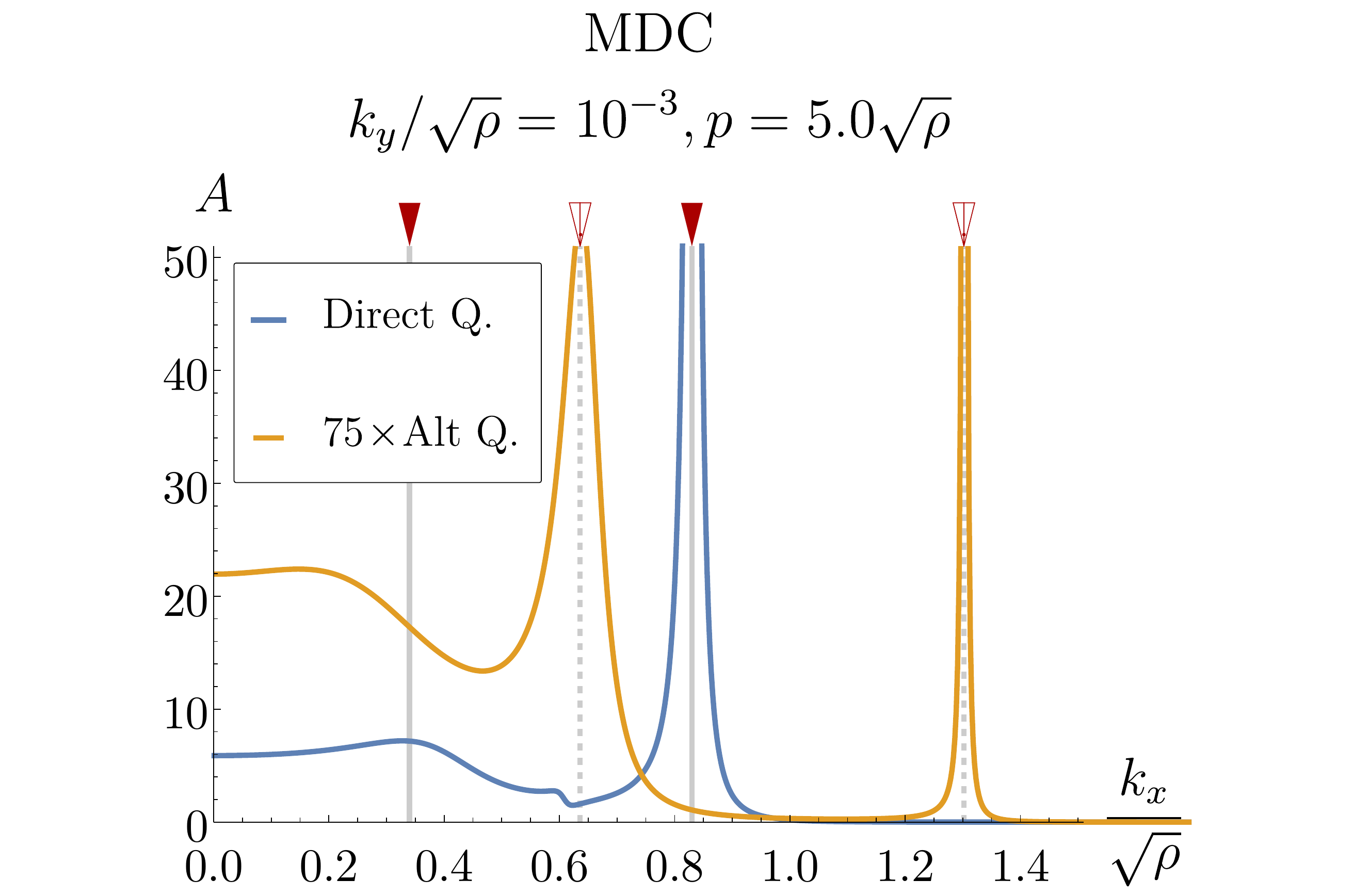} \\[1cm]
\hspace{-0.5cm}
\includegraphics[height=\myhight cm]{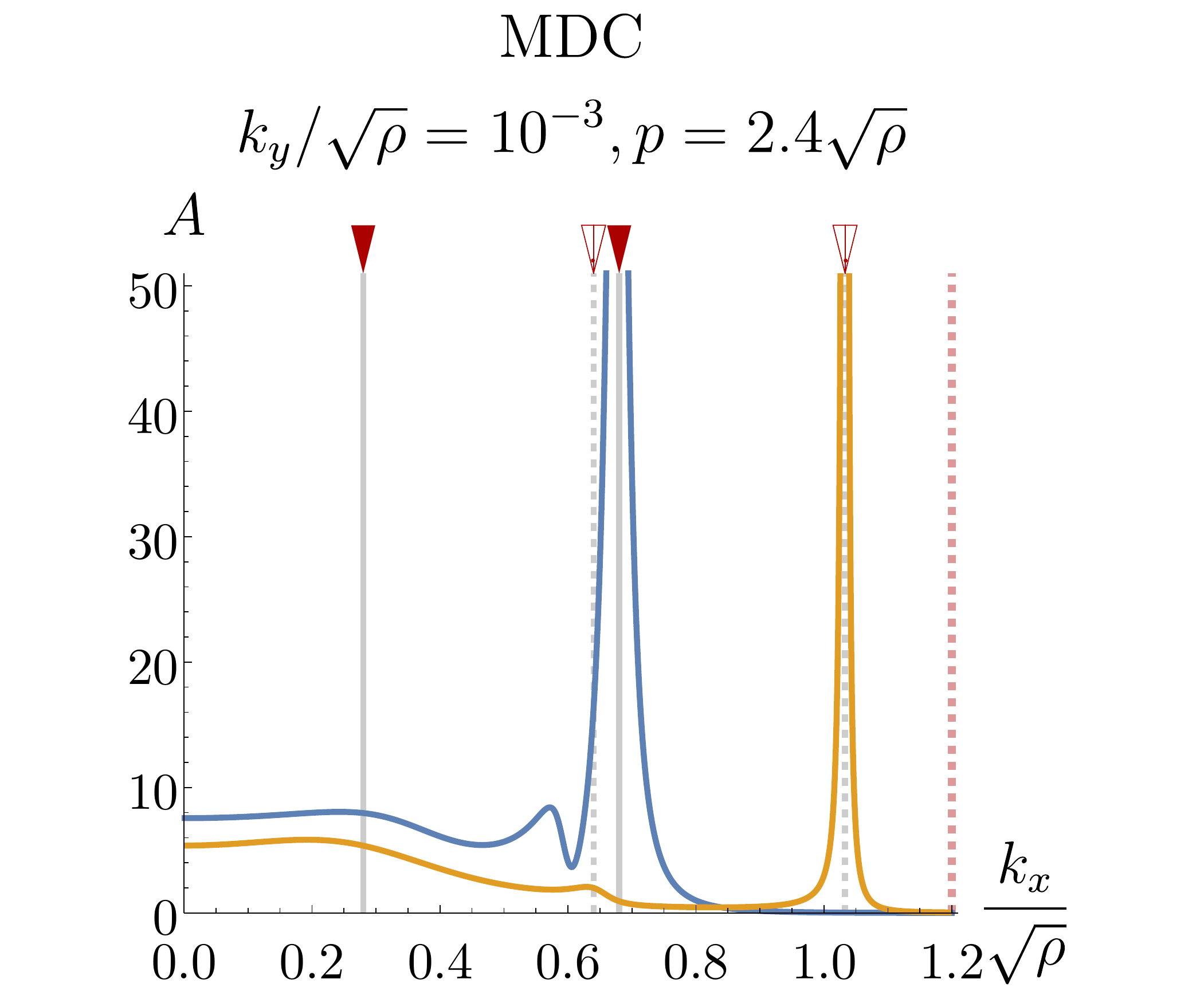} 
\hspace{1cm}
\includegraphics[height=\myhight cm]{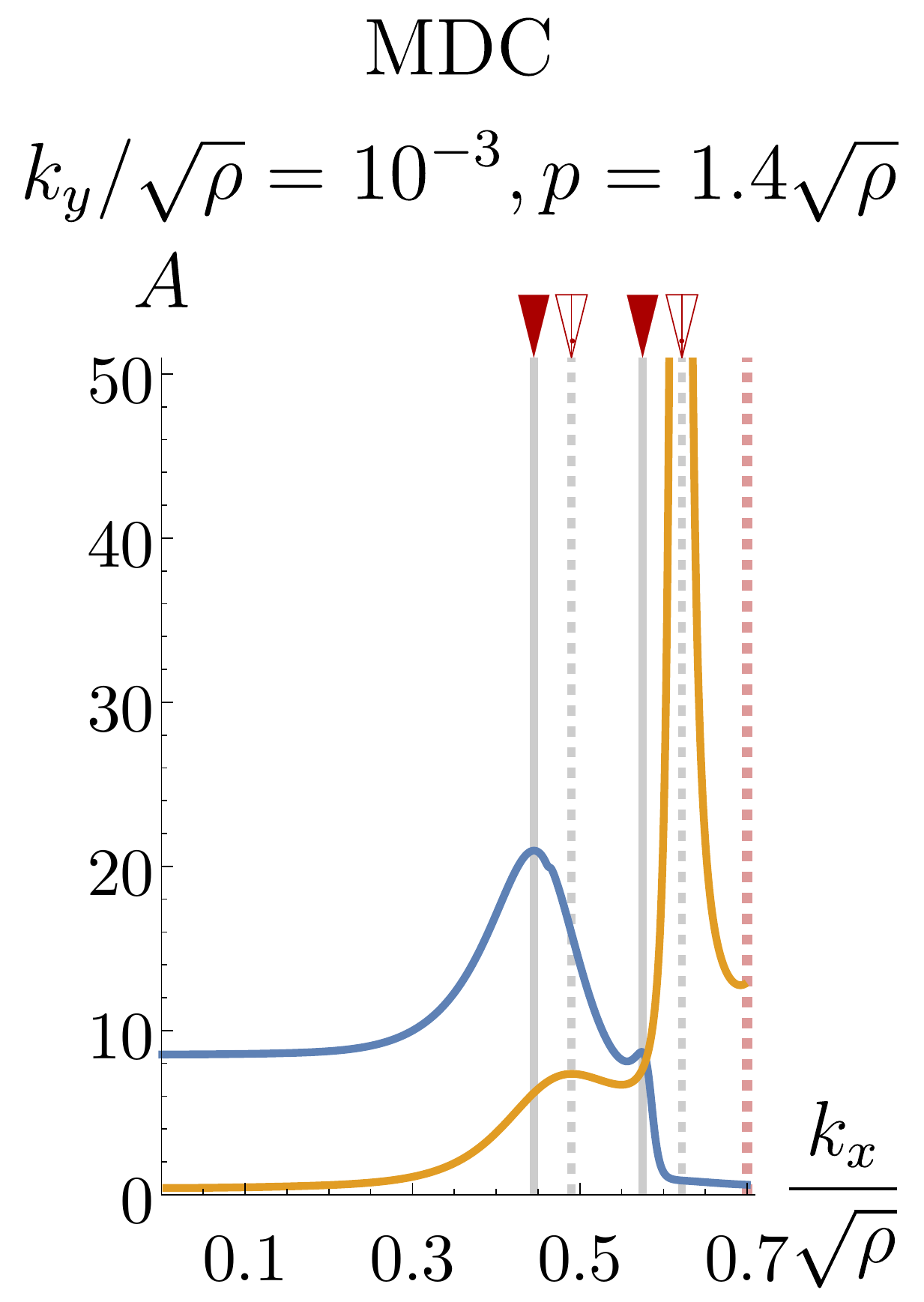} 
\caption{\label{fig:MDC}\textbf{Momentum distribution curves in strong unidirectional holographic lattices ($\lambda = 8.0$) in direct and alternate quantizations.}\\
At large lattice momentum, when the BZ boundary is far away from the FS, the the peaks in direct and alternate quantization alternate along the MDCs (c.f. Fig.\,\ref{fig:RN_fermions}). The positions of Direct Q peaks and Alt Q dips are correlated. When the BZ boundary is brought closer, the peaks and zeroes seem to absorb each other. For $p = 1.4\sqrt{\rho}$, the quasiparticle peak is eaten by the zero completely. 
}
\end{figure}

We can expose the ``zero-eats-pole'' effect in more detail by analysing cuts along the $k_x$-axis of the spectral density. These Momentum Distribution Curves (MDC) are shown in Fig.\,\ref{fig:MDC}, where we plot the Direct and Alternative quantizations, in exactly the same fashion as we did in Fig. \ref{fig:RN_fermions}. A clear depletion is observed in the Direct Q MDC corresponding to the secondary FS peak in the Alt.\,Q scheme. As the size of the Brillouin zone (red dashed grid line) is decreased as shown in bottom left plot of Fig. \ref{fig:MDC}, this depletion approaches the peak shown by filled red pointer and absorbs it completely at $p=1.4 \sqrt{\rho}$ as shown in the bottom right plot of Fig. \ref{fig:MDC}.
Therefore, we conclude that for strong lattices, the Fermi surfaces in both direct and alternative quantizations are deformed in such a way that the poles and zeros overlap and cancel each other.

\begin{figure}[ht]
\center
\newcommand\myhight{7}
\includegraphics[height=\myhight cm]{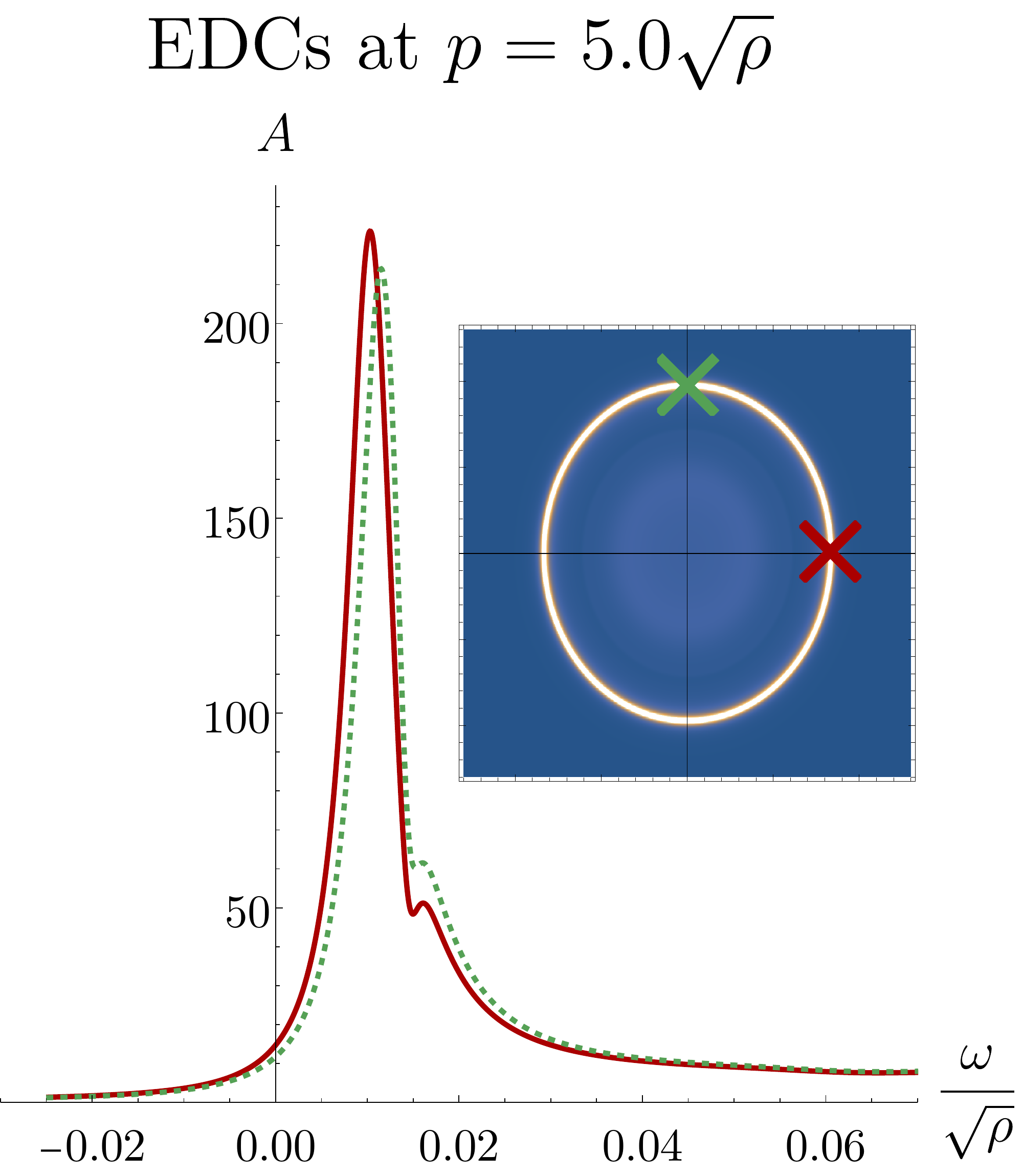} \\[2.5cm]
\includegraphics[height=\myhight cm]{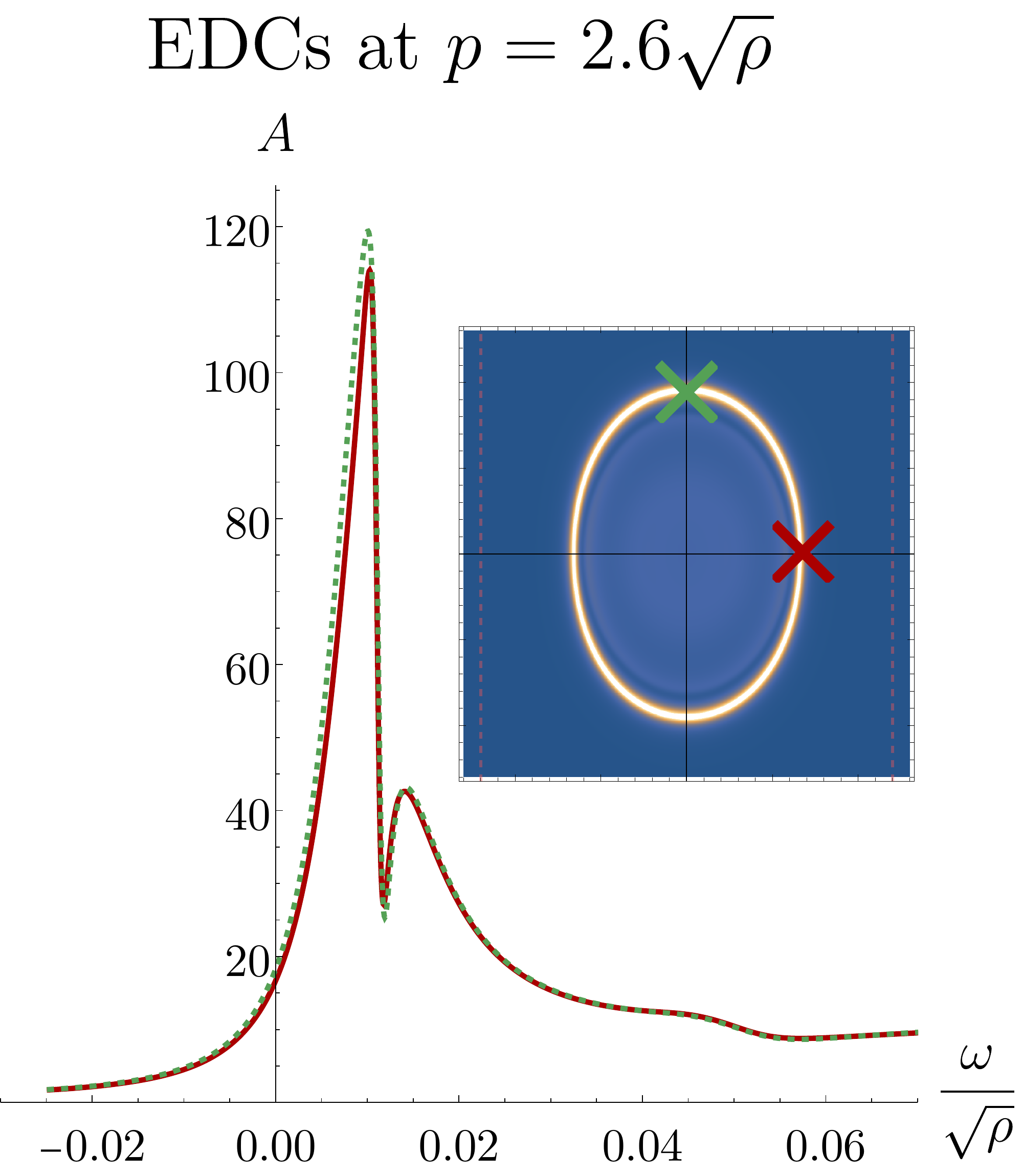} 
\includegraphics[height=\myhight cm]{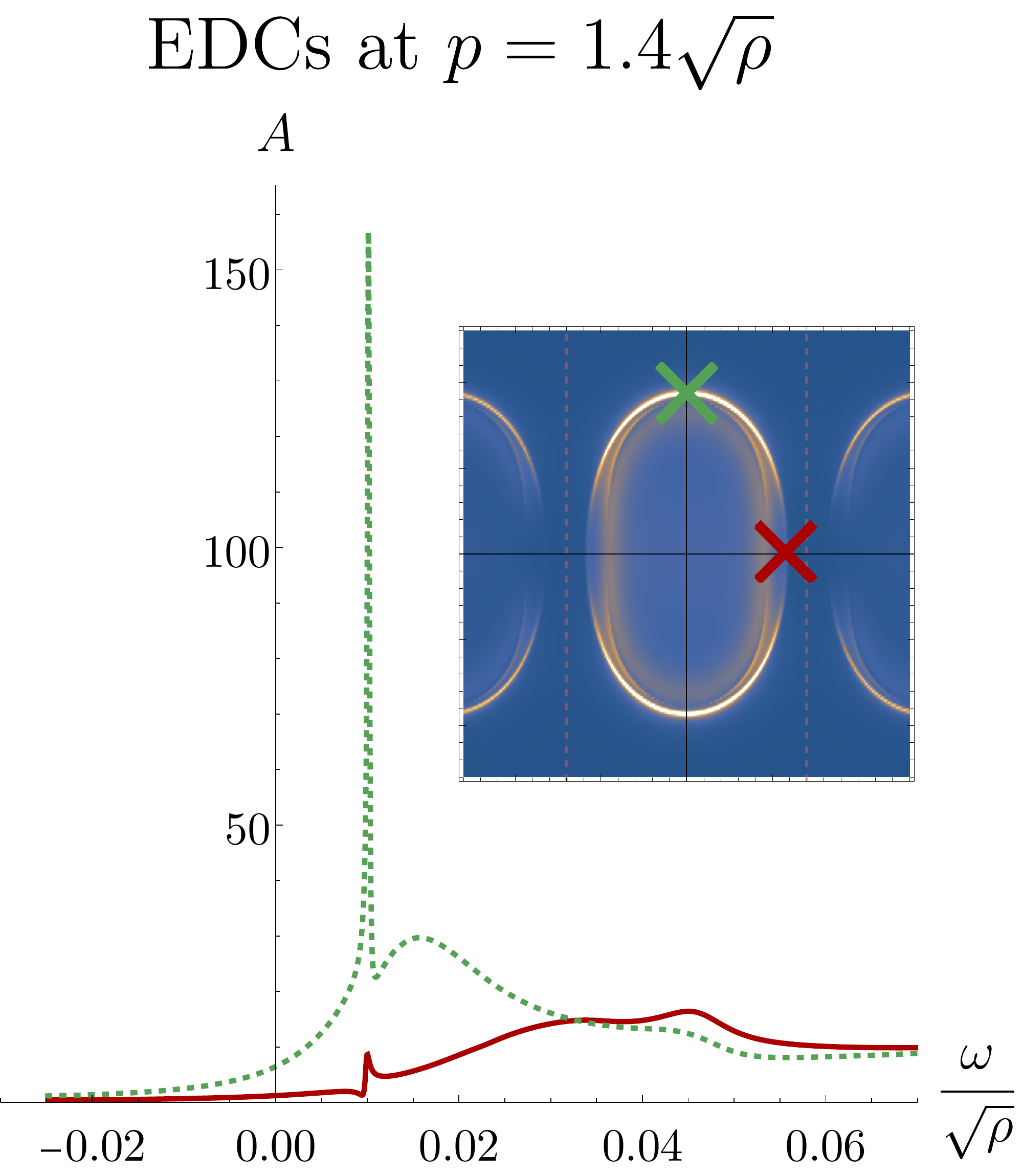} 
\caption{\label{fig:EDCs}\textbf{Energy distribution curves in for strong unidirectional holographic lattices ($\lambda = 8.0$) in direct quantization near and far from the Umklapp surfaces.}\\
When the BZ boundaries are far away, as in the top picture, the EDCs look very similar: they both show a well-defined, sharp peak. The dip on the right shoulder of the peak corresponds to the alternate quantization zero. When the BZ boundary is brought closer, first some asymmetry in the sharpness of the peaks appears. When the BZ boundaries are brought very close, the peak for the EDC near the BZ boundary gets destroyed by the zero completely.
}
\end{figure}

This destruction has a dramatic effect on the very existence of the quasiparticle excitations in part of the kinematic region. This is best seen on Energy Distribution Curves (EDCs), which we show on Fig.\,\ref{fig:EDCs}. Here, we plot the frequency dependence of the spectral weight at two perpendicular directions in the Fermi surface: $k_x = k_f^{(x)}, k_y= 0$ and  $k_x = 0, k_y = k_f^{(y)}$.
When the BZ boundary is far away (top panel of Fig.\,\ref{fig:EDCs}), the EDCs are practically identical in both directions, confirming that the FS is almost isotropic. In the bottom left panel, the anisotropy is now manifest, since the FS is deformed by the lattice. Finally, the most drastic effect is seen on the bottom right panel of Fig.\,\ref{fig:EDCs}. On the one hand, in the $k_y$-direction, there is still a sharp peak corresponding to a quasiparticle excitation. However, the spectral density in the $k_x$ direction (red line) is totally incoherent -- there is no excitation with definite energy which  propagates along the $k_x$-direction.

\section{\label{sec:conclusion}Discussion}
In this work we have studied the fermionic spectral function in a holographic model with periodic ionic lattices. Our most important finding is a novel phenomenon which appears as a destruction of coherent spectral weight peaks of the Fermi surface along the  directions of the lattice vector. We have shown that the origin of this phenomenon is the interaction between poles and zeros of the fermionic Green's function. More precisely, due to non-linear effects from the strong background lattice potential, these are pushed together and suppress each other. What is tantalizing, is that the patterns of the spectral density observed here look very similar to the results of ARPES experiments in strongly correlated materials in which nodal-antinodal dichotomy and Fermi arcs are observed. There one also observes the destruction of the coherent spectral weight in certain directions. It is therefore very interesting to further investigate  whether our results can clarify these unconventional phenomena. 

A warning is in order however; one should handle the results of holographic models with great care.
As we discussed in Sec.\,\ref{sec:holographicFS}, the presence of zeros in our holographic treatment follows from the simple logic relying on the existence of two possible choices for the dual CFT operators corresponding to each quantization, and the existence of a Fermi surface in the alternate quantization.
However, as we noted on Fig.\,\ref{fig:RN_fermions}, when only one quantization is allowed, zeros are still present and perhaps their existence is a more fundamental feature, which is yet to be explored. The fact that the zeros and (the multiple) poles of the fermionic Green's function appear in simple models as alternating series suggests that, as we drive some parameter (chemical potential or temperature) to zero, they might coalesce and form a branch cut in the complex plane. This signals the truly unparticle shape of the Green's function characteristic of the ultraviolet CFT of the holographic duality.
This idea is rather at odds with the physics of experimentally realized strange metals, because the UV CFT of a holographic model is clearly not the UV theory governing the behaviour of electrons in real materials. Moreover, the multiple pole features of the fermionic response in the holographic model arguably rely on the large-$N$ approximation, which may or may not be efficient in reality.

However, one can regard our results with a larger perspective;  we observe that the destruction between  zeros and poles requires the two fundamental ingredients: the very existence of zeros in the fermionic Green's function and  strong lattice potentials, which influences the position of poles and zeros by bringing them closer together so that eventually they annihilate each other. Even though the understood origin of the zeros in our particular holographic model cannot be directly mapped to known condensed matter systems, there are models in condensed matter theory, which display similar features, in particular Mott insulators \cite{phillips2006mottness,dzyaloshinskii2003some}. Our finding suggests that, if upon doping the insulating Mott zeros of the Green's function remain present, their proximity to the quasi-particle poles may indeed be the origin of this spectacular experimental phenomenon. Doped Mott insulators are exactly the systems where the real Fermi arcs are observed. In this regard holographic studies certainly confirm its use as a convenient theoretical laboratory for studying the phenomenology of the strongly correlated systems which goes far beyond the applicability of the Fermi liquid theory.

\acknowledgments{
We thank 
Nikolay Gnezdilov, Andrei Bagrov, Askar Iliasov, Li Li, Sera Cremonini, Saso Grozdanov and Richard Davison for fruitful discussions.
We appreciate the  useful comments we've got from Michael Norman, Philip Phillips, Robert Leigh and Blaise Gouteraux.
We acknowledge the contribution of Stefan Vandoren and Simon Gentle who participated at the early stages of this project.

This work is a part of the Strange metal consortium, funded by Foundation for Research into Fundamental Matter (FOM) in the Netherlands.

This research was supported in part by Koenraad Schalm's VICI award of the Netherlands Organization for Scientific Research (NWO), by the Netherlands Organization for Scientific Research/Ministry of Science and Education (NWO/OCW), and by the Foundation for Research into Fundamental Matter (FOM).

A.K. appreciates the opportunity to deliver and discuss the preliminary results of this study, during the conference ``Gauge/Gravity Duality 2018'' in the University of W\"urzburg and the workshop ``Bringing Holography to the Lab'' in Lorentz Center, Leiden.   
A.K. acknowledges the hospitality of CPHT group of \'Ecole Polytechnique, Paris and the useful feedback they provided. 

The numerical calculations were performed on the Maris Cluster of Instituut Lorentz.
}

\newpage

\appendix

\section{\label{app:RN_fermions}Fermionic equations of motion in the RN background}

Here, we discuss in  detail the algorithm to numerically solve the Dirac equation in the RN background \eqref{equ:Dirac_RN_expanded} for two cases depending on the bulk fermion mass:
\begin{itemize}
\item $m=\tfrac{1}{4}$: The case $|m|<1/2$, when the alternative quantization is allowed.
\item $m=\tfrac{3}{4}$: The case $|m|>1/2$, when the alternatively quantized CFT on the boundary side is ill-defined (non-renormalizable) \cite{Cubrovic:2009ye,Laia:2011wf}.
\end{itemize}
In both cases on top of the direct approach we will also formally evaluate the response in the alternative quantization (setting $m \rar -m$), which can be done in the bulk even in case of $m=3/4$, despite the fact that the boundary dual theory is ill defined.

In order to eliminate divergences at the UV boundary \eqref{eq:boundary_expansion} and the oscillations due to the ingoing boundary conditions at the horizon \eqref{eq:infalling}, we redefine the components of the wave function in \eqref{equ:Dirac_RN_expanded} 
\begin{equation}
\label{equ:rescale}
\zeta(z) \rar z^{-m} (1 - z)^{-i \frac{\omega}{4 \pi T}} \xi(z),
\end{equation}
and solve for $\xi(z) \equiv (\xi_r,\xi_s)^T$ in what follows. 

For our choices of the bulk mass,  the rescaling of Eq.  \eqref{equ:rescale} leads to one of the components approaching a constant at the boundary $z\rar 0$, and the other component scaling as a positive half-integer power of $z$. The terms in the equations of motion contain only integer powers of $z$ at the boundary, therefore one can expand the solution as series (see \eqref{eq:boundary_expansion})
\begin{align}
\label{equ:boundary_series}
z& \rar 0
&m &= \frac{1}{4}: &
\begin{cases}  
\xi_r &= b z^{1/2} \left(1 + \mathfrak{b}^1_r z + \dots \right) + a z \left(\mathfrak{a}^0_r + \mathfrak{a}^1_r z + \dots \right) \\
 \xi_s &= a \left(1 + \mathfrak{a}^1_s z + \dots \right)   + b z^{3/2} \left(\mathfrak{b}^0_s + \mathfrak{b}^1_s z + \dots \right),
\end{cases}\\
& &m &= \frac{3}{4}: &
\begin{cases}  
\xi_r &= b z^{3/2} \left(1 + \mathfrak{b}^1_r z + \dots \right) + a z \left(\mathfrak{a}^0_r + \mathfrak{a}^1_r z + \dots \right)  \\
\xi_s &= a \left(1 + \mathfrak{a}^1_s z + \dots \right) + b z^{5/2} \left(\mathfrak{b}^0_s + \mathfrak{b}^1_s z + \dots \right),
\end{cases}
\end{align}
where the coefficients $\mathfrak{a}_a^l, \mathfrak{b}_a^l$ are fixed by the near boundary expansions of the equations of motion while $a,b$ are the integration constants describing the source and response, respectively. Importantly, we see that because the expansion is in half-integer powers of $z$, the  series proportional to $a$ and $b$ never overlap and therefore there are no logarithmic terms appearing in the expansion \cite{Skenderis:2002wp,Bianchi:2001kw}. 

Near the horizon, both spinor components approach  constants for any value of mass. However, the equation coefficients include the factors of $\sqrt{f} \sim \sqrt{1-z}$, therefore the expansion near horizon goes in half-integer powers of $(1-z)$
\begin{align}
\label{eq:horizon_series}
z &\rar 1: &
\xi_r &= \ \ \, \ h (1 + \mathfrak{h}_1^1 \sqrt{1-z} + \mathfrak{h}_1^2 (1-z) \dots) \\
\notag
& &
\xi_s &= - i h (1 + \mathfrak{h}_2^1 \sqrt{1-z} + \mathfrak{h}_2^2 (1-z)  + \dots),
\end{align}
where $h$ is an arbitrary integration constant and the coefficients $\mathfrak{h}_a^l$, including the ``$-i$'' factor in front of $\xi_s$ come from the expansion of the equations of motion near horizon. 

In order to render the solutions regular on both ends of the integration interval it is therefore convenient to rescale the coordinate as well. More specifically, we define
\begin{equation}
\label{eq:coordinate_rescale}
z \equiv \left(1 - (1-r)^2 \right)^2, \qquad r\Big|_{z\rar0} \sim \sqrt{z}/2, \qquad (1-r)\Big|_{z\rar 1} \sim \sqrt{(z-1)/2}.
\end{equation}

With these modifications, the problem reduces to  solving two coupled first-order linear ODEs \eqref{equ:Dirac_RN_expanded} on the interval $r \in [0,1]$ subject to the boundary conditions \eqref{equ:boundary_series}, \eqref{eq:horizon_series}. There are 3 integration constants $a,b$ and $h$, one of which can be set to unity due to the linearity. The remaining two are fixed by the two boundary conditions giving  a unique solution for fixed $\omega, k_x, k_y$. We use the numerical shooting method to obtain this solution. We shoot from  both ends of the interval, using the expansion series \eqref{equ:boundary_series} and \eqref{eq:horizon_series} as the initial conditions, and  look for the values of free constants $b,h$ for which the solutions match in the arbitrary point in the interior of the interval. The advantage of this method is that we have a direct control over the response constant $b$ and do not  need to extract it from subleading terms in the near boundary expansions of the solution as it is usually done when shooting from horizon only.

\section{\label{app:numerics}Numerical calculus and precision control for gravity background}
In order to obtain the non-homogeneous gravitational background with a periodic boundary condition for the gauge field \eqref{eq:mux_hologrpahy}, we have to solve the full set of the Einstein equations. 
These follow from Eq. \eqref{equ:Einstein_Maxwell_action}, which gives a set of 6 coupled non-linear partial differential equations (PDEs) in coordinates $(x,z)$, which are not elliptic.
In practice we solve for functions $\hat{Q}_{\mu \nu}, \hat{A}_\mu$ defined as
\begin{align}  
A_t &= \bar{\mu} (1-z) \hat{A}_t, & \mathcal{T}^2 &= 1 + z \hat{Q}_{tt}, & \mathcal{X}^2 &= 1 + z \hat{Q}_{xx} \\
 \mathcal{Y}^2 &= 1 + z \hat{Q}_{yy}, &  \mathcal{Z}^2 &= 1 + z \hat{Q}_{zz}, & Q_{xz} & = z \hat{Q}_{xz},  
\end{align}
such that the RN black hole solution corresponds to the choice $\hat{Q}_{\mu \nu}(x,z) = 0, \hat{A}_t(x,z) = 1$.

We use the DeTurck trick, outlined in \cite{Wiseman:2011by,Headrick:2009pv,Adam:2011dn} and used for this type of holographic models in \cite{Horowitz:2012gs,Rangamani:2015hka}. It allows to recast the Einstein's equations as a boundary value problem for a set of nonlinear elliptic equations. 
We choose the RN black hole as a reference metric, which allows us to use for the non-homogeneous background the same expression of temperature as a function of chemical potential as that of the RN black hole \eqref{equ:temperature}. 

The boundary conditions in $x$-direction are periodic, as  dictated by the symmetry of the background lattice potential. 
It is convenient to rescale the spatial coordinate $x \rar (2 \pi/p) \hat{x}$ in order to fix the integration interval to unity: $\hat{x} \in [0,1)$. 
At the $z \rar 0$ boundary we require the metric to be asymptotically AdS with a nontrivial inhomogeneous source for the gauge field (c.f. \eqref{eq:mux_hologrpahy})
\begin{equation}
z \rar 0: \qquad \hat{Q}_{\mu \nu}(x,0) = 0, \qquad \hat{A}_t(x,0) = 1 + \lambda \cos(2 \pi \hat{x}).
\end{equation}
The horizon boundary conditions follow from the asymptotic expansion of the Einstein equations near the horizon $f(1) = 0$. We expand the unknown functions in the Taylor series down to the first derivative order and substitute these expansions into the Einstein equations. This gives us 5 equations in the subleading order which relate the derivatives of the fields to their horizon values, i.e. generalized Robin boundary conditions. On top of that we get one algebraic equation $\hat{Q}_{tt}(x,1) = \hat{Q}_{zz}(x,1)$ at the leading order, which guarantees the solution to be static \cite{Horowitz:2012gs,Donos:2014yya,Rangamani:2015hka}.

The numeric algorithm follows closely the treatment of our earlier works \cite{Mott,Krikun:2017cyw,Krikun:2018ufr} with several technical improvements.
%
We used a finite difference discretization of the equations on a homogeneously spaced grid and realized a Newton-Raphson procedure to solve the nonlinear boundary value problem. While supplementary procedures were implemented, including a relaxation scheme with Orszag regularization, no other methods matched the speed and accuracy of the direct Newton-Raphson method \cite{boyd2001chebyshev,trefethen2000spectral}, see also \cite{Krikun:2018ufr}. \\

We ran a comprehensive series of precision and accuracy tests to make a judicious choice of computational parameters. We employed the norm of the DeTurck vector and the value of the trace of the Einstein tensor as measures of convergence. The thermodynamic potential was used as a physical observable, from which we could estimate the relative precision of our calculations.
From this analysis, we were able to conclude that the configuration which gives us the best balance between speed and accuracy of computations for both the backgrounds and the fermions is a grid of $n_x \times n_y = 34\times80$ points, where we use 8th-order accuracy central finite difference derivatives in our finite difference scheme. This achieved relative errors of less than $10^{-5}$ in the thermodynamic potential, while still allowing for the fermionic equations to be computed quickly and accurately enough. In order to make sensible comparisons, 
we work in the canonical ensemble with  a constant charge density. Since this is not a parameter in our equations, but rather an observable we can only extract after solving the background, we employed a root-finding algorithm with a relative tolerance of $10^{-3}$ to fix the charge density. 

We implement our numerical routines in Python 3.6 using a set of packages that can be found in the standard SciPy stack~\cite{scipy}. Under the hood, these use the SuperLU~\cite{li05} package for solving the sparse linear equations. 
We have performed several cross checks with a similar code in Mathematica \cite{Mathematica10}, which was used earlier in \cite{Andrade:2017leb,Mott,Krikun:2017cyw}, in order to prove the reliability of the package. The code was designed such that it could run a large number of small instances in parallel on any number of machines. This technique was suitable for running on the Lorentz Institute \emph{Maris} cluster

\section{\label{app:fermi_numerics}Numerical calculus for Dirac equation}
There are few subtleties which one encounters when solving the Dirac equation on the non-homogeneous background \eqref{equ:Dirac_lattice} numerically. First, in order to make the fermionic solutions regular on both horizon and asymptotic boundary we perform the same set of redefinitions of the wave-functions \eqref{equ:rescale} and holographic coordinate \eqref{eq:coordinate_rescale} as it was done in the homogeneous case discussed in Appendix\,\ref{app:RN_fermions}. 

Moreover, there are several strategies which one can use in order to solve the set of first order differential equations numerically. One of them is to solve a Cauchy problem, integrate the equations starting from horizon and read off the boundary asymptotes. This is somewhat similar to the shooting method we discussed above and it was used in the early work \cite{Ling:2013aya}. However, when dealing with  PDEs, one has to consider solving for all possible Fourier modes on the horizon and tracing out all the components of the response matrix $\mathcal{S}_{nl}$ \eqref{equ:S_matrix} before one gets access to the single $\mathcal{S}_{00}$ component, which we are mostly interested in. Therefore we find this approach not suitable.
Instead, we consider the boundary value problem, imposing the (position dependent) infalling boundary conditions \eqref{eq:horizon_series} at the horizon as well as setting the source $a(x)$ on the asymptotic boundary $z\to0$ to a desired Fourier mode. In this way, setting for instance $a(x) = e^{i p l}$ with fixed $l$ and measuring the full profile of $b(x) = \sum_n e^{i p n}$, we get the information on all components of the response matrix in a row $\mathcal{S}_{ln}, n \in [-N/2,N/2]$, where $N$ is the size of the grid in the $x$-direction. This boundary value approach is more useful since we can obtain the desired value $\mathcal{S}_{00}$ 
in a single iteration by considering simply $a(x) = 1$ as  boundary condition. 

However, a problem arises when one tries to implement the boundary value PDE solving code in the case of a Dirac equation. 
The boundary value problem requires setting the boundary conditions for each field on both ends of the integration interval. 
For a first order differential equation, like the Dirac equation, this over determines the problem. 
We avoid this obstacle by formally setting a trivial boundary conditions of the form `$0=0$' for half of the fields on the boundary. 
Indeed, after performing the rescaling of the fields \eqref{equ:rescale} we guarantee that the sub-leading spinor component will behave as $\xi_r \sim z^{2 m}$ at the boundary. Therefore, setting $\xi_r(0) = 0$ does not impose any extra constraint in the problem. 
This precise trick does not work however in case of alternative quantization, when we formally take $m < 0$ and the ``response'' branch $\sim z^{2 m}$ diverges on the boundary. In the particular case of $m = - 1/4$ studied  here,  this divergence is mild enough and we simply get rid of it by further redefining the ``response'' fermion component with an extra factor of $z$. More precisely, in case of $m=-1/4$ instead of \eqref{equ:rescale} we use $\zeta_r \rar z^{-m - 1} (1-z)^{-i \frac{\omega}{4 \pi T}} \xi_r$. This changes the near boundary behavior of the $\xi_r$ component to $\xi_r \sim z^{1/2}$. As before, the Dirichlet boundary condition $\xi_r = 0$ is trivial and does not over constrain the problem.

We use a similar logic on the horizon; in addition to the leading behaviour $\xi_r (x)= i \xi_s (x)$, Eq.  \eqref{eq:horizon_series}, following from the ingoing boundary condition, we include the sub-leading terms in the expansion of the equations of motion,. These, relate the derivatives of the functions to their boundary values.
Since these relations are obtained from the equations of motion themselves, they do  not introduce extra constraints and we are left with the correct amount  of boundary conditions. 

Another problem with the first order differential equations is that the matrix which represents the discretized problem on a lattice does not have a positively defined spectrum of eigenvalues typical of elliptic problems. 
Therefore, one can not rely on the iterative methods, because  these are not guaranteed to converge. In our case, the Dirac equation is linear and one does not require to use iterations of any kind: the problem is solved ``in one shot'' by inverting the master matrix once. 
Even though we managed to make use of this approach, it may not always be applicable, especially in  cases where the pseudospectral collocation is used and hence the matrix is dense, or simply if the grid is dense and inversion of the huge matrix is not feasible. 
In this case one can improve the situation by substituting the first order equations \eqref{equ:Dirac_lattice} with the second order elliptic ones.
Indeed, if we represent the equations \eqref{equ:Dirac_lattice} for $\xi^{\ua}$ and their P-transformed counterparts for $\xi^{\da}$ as
\begin{equation}
\mathrm{Dirac}^{\ua} \equiv \mathcal{D}_{x} [\xi^{\ua} ] + \mathcal{K}_{y} \xi^{\da}, \qquad  \mathrm{Dirac}^{\da} \equiv \mathcal{D}_{-x} [\xi^{\da} ] - \mathcal{K}_{y} \xi^{\ua}\,,
\end{equation}
we can apply to one of the equations the differential operator of the other equation, and vice versa, to get the following second order system
\begin{align}
\label{equ:elliptic_Dirac}
(\mathrm{Dirac}^{\ua})^2 &= \mathcal{D}_{-x} [\mathcal{D}_{x} [\xi^{\ua} ]] + \mathcal{D}_{-x} [\mathcal{K}_{y}] \xi^{\da} + \mathcal{K}_{y}^2 \xi^{\ua}, \\
\notag
(\mathrm{Dirac}^{\da})^2 &= \mathcal{D}_{x} [\mathcal{D}_{-x} [\xi^{\da} ]] - \mathcal{D}_{x} [\mathcal{K}_{y}] \xi^{\ua} + \mathcal{K}_{y}^2 \xi^{\da}\,.
\end{align}
In this form, the second order differential operator for each  spinor component is the Laplacian in curved space. This representation allows us to use the full power of the iterative numerical techniques designed for elliptic equations. 
We have observed that the direct inversion of the master matrix, similar to the one we used in the first order case, becomes numerically less demanding due to improved features of the differential operators. 
When using the second order equations \eqref{equ:elliptic_Dirac} one has to take care that no ghost solutions are obtained which do not solve the original problem. 
We can guarantee this by using the expanded first order Dirac equations as the boundary condition on the horizon. 
Unlike the first order case, these boundary conditions are not trivial, but they rather impose the constraint that the first order equations are satisfied at the horizon, and this constraint is further propagated in the bulk by the second order system. 
We checked in our numerical calculations that the two approaches, with the first and the second order differential equations, give identical results and this cross-check serves as a good confirmation of the validity of our numerical treatment.

\section{\label{app:BZ}Green's function in the Bloch momentum representation}

Here, we  study spatial features of the Green's function in the Bloch momentum representation and the way the poles in the alternative quantization manifest themselves as zeroes in direct quantization. We focus on the response matrix \eqref{equ:lattice_S_matrix}. 
In practice, by representing the Dirac equation as a boundary value problem, as outlined in Appendix\,\ref{app:fermi_numerics}, we have a direct control over $a(x)$ and can, for instance, source any given harmonic mode. In most cases we just switch on $a_0 = 1$, corresponding to the constant source and obtain $b(x)$ as the series
\begin{equation}
b^{(0)}(x) = \sum_n e^{i n p x} \mathcal{S}_{n 0}\,.
\end{equation}
Then, we extract the $b_0$ homogeneous component, which gives  $\mathcal{S}_{00} = b_0/1$ (note that we normalized $a_0 = 1$ here). 

\begin{figure}
\center
\begin{minipage}{0.49\linewidth}
\center
\includegraphics[height=8cm]{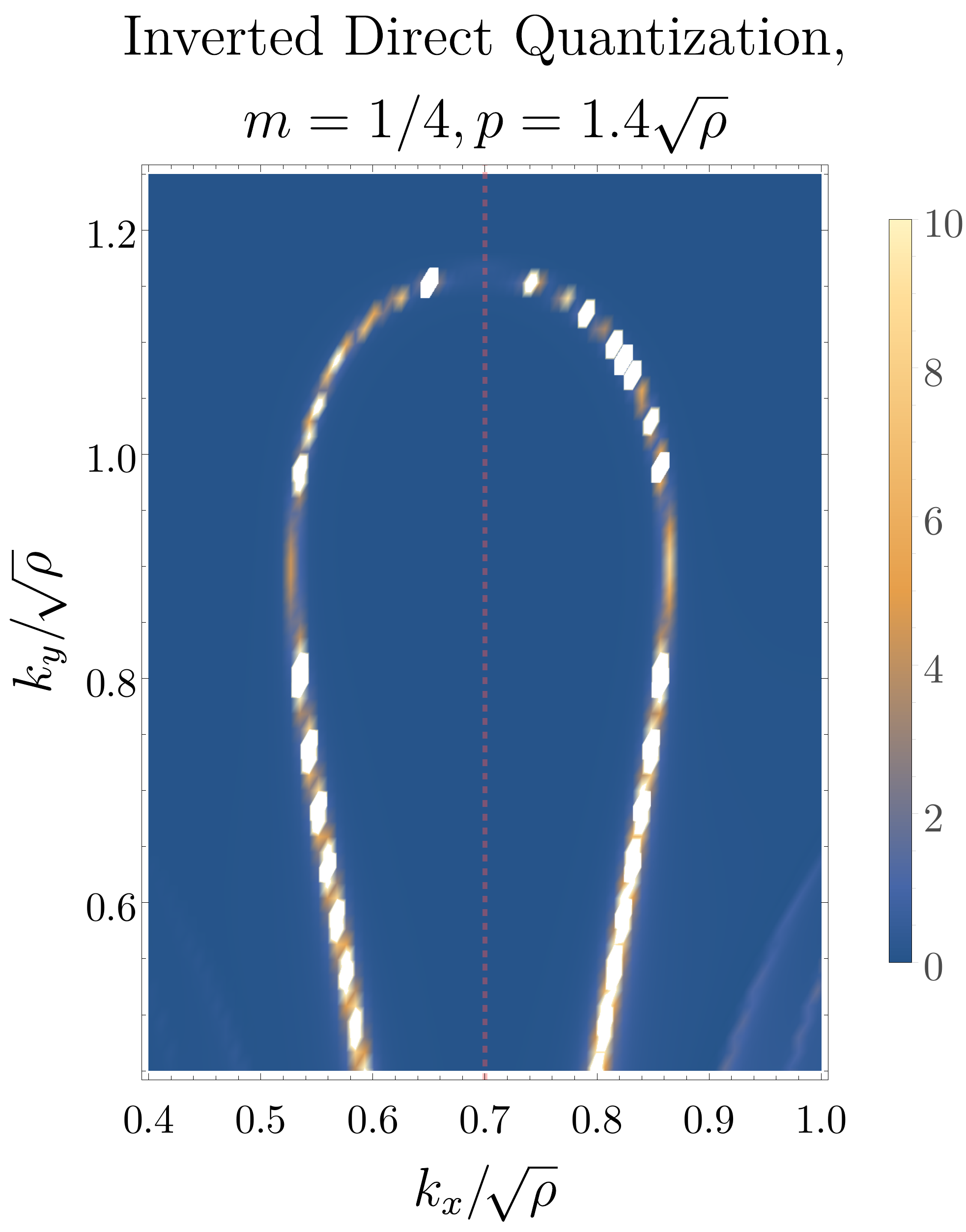}
\end{minipage}
\begin{minipage}{0.49\linewidth}
\center
\includegraphics[height=8cm]{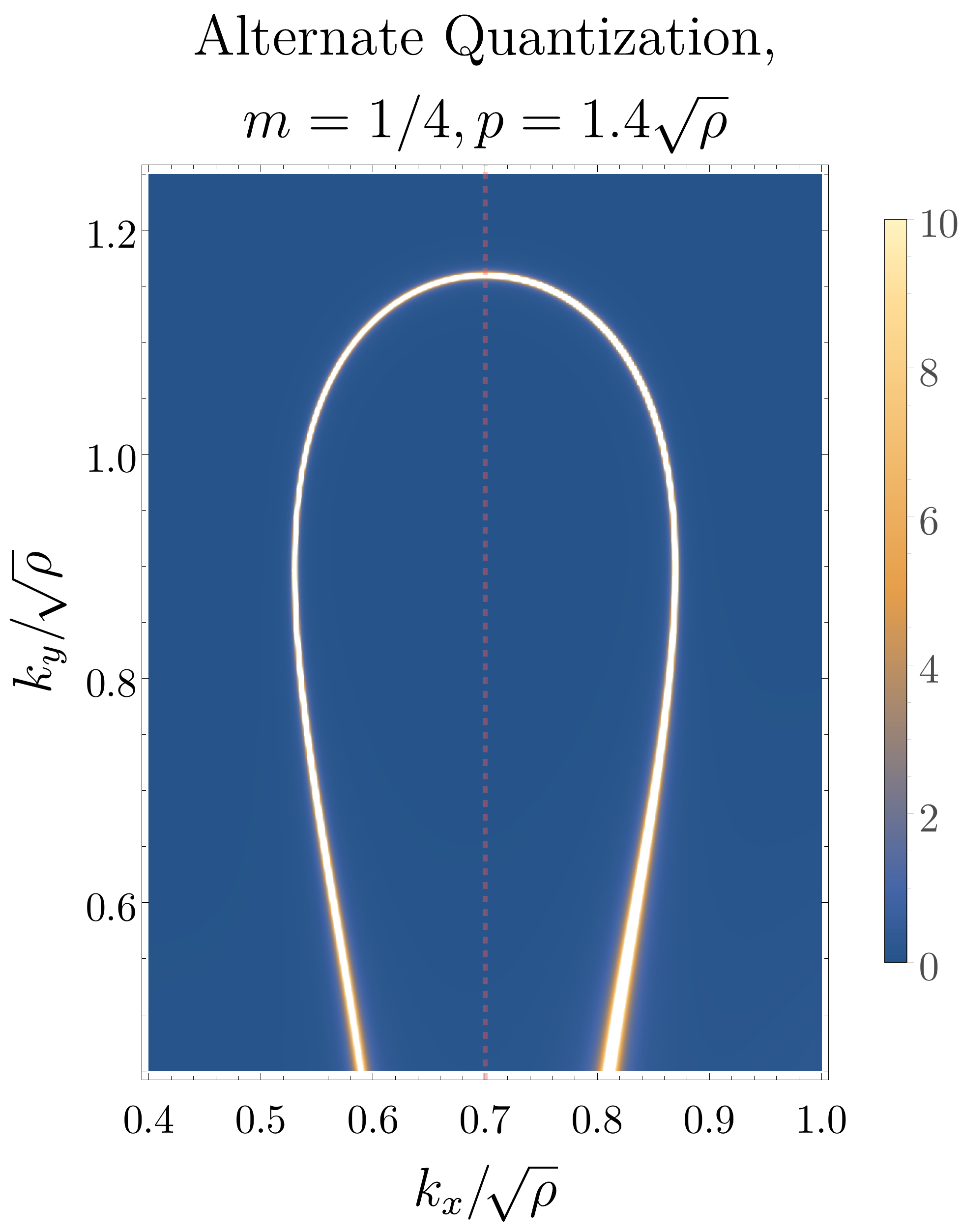}
\end{minipage}
\caption{\label{fig:altQ_two_treatments} \textbf{Comparison of the two treatments of the alternative quantization picture}
The comparison between computation in alternate quantization and doing the inversion procedure from the direct quantization data. The main features (sharpness, shape, and the reduction of spectral weight near $(k_x, k_y) = (0.65, 0.40)\sqrt{\rho}$) are demonstrably reproduced. The inversion procedure does suffer from limited numerical precision, as the contributions from all the different Green's function components range over many orders of magnitude.  The non-smoothness of the left-hand side arises because the width of the FS is smaller than the density of data points, causing the computation to not quite hit the peak. This can be improved in principle, however due to the prohibitive computational cost of this procedure we don't show it here.
}
\end{figure}

In principle, we can go further and study a full rectangular sector of a (formally infinite) $\mathcal{S}$-matrix.
In order to do so, we consider several harmonic sources $a^{(l)}(x) = e^{i l p x}$ and evaluate the responses for these cases
\begin{equation}
b^{(l)}(x) = \sum_n e^{i n p x} \mathcal{S}_{n l}.   
\end{equation}
It is now clear that by extracting  $N$ Fourier modes of the response profiles $b^{(l)}(x)$ for a set of $L$ harmonic sources we get access to the full $N\times L$ patch of the $\mathcal{S}$-matrix.
Formally, in order to treat the alternative quantization as  exchanging the of roles between $b(x)$ and $a(x)$ and obtain the $\tilde{\mathcal{S}}_{00}$ component of the Alt.\,Q response matrix, we have to guess what kind of the boundary ``response'' condition $a(x)$ would lead to a constant ``source'' $b(x) = b_0$. In other words, our goal in this case is to find a set of coefficients $a_l$, such that
\begin{equation}
\sum_l a_l b^{(l)}(x) = \sum_{l,n} \mathcal{S}_{n l} a_l e^{i n p x} = b_0.
\end{equation}
In a vector notation this has a simple form
\begin{equation}
\mathcal{S} \cdot \vec{a} = b_0 \vec{e}_0, 
\end{equation}
where $\vec{a} = (\dots, a_{-1}, a_0, a_1, \dots)$ and $\vec{e}_n$ is the unit basis vector with the only nonzero component at $n$-th position. Clearly, the result is
\begin{equation}
 \vec{a} = b_0 \mathcal{S}^{-1} \cdot \vec{e}_0
\end{equation}
And after taking the $a_0$ component we arrive at the expression: $\tilde{\mathcal{S}}_{00} \equiv b_0/a_0 = (\mathcal{S}^{-1})_{00}$, from which the equation \eqref{equ:lattice_alt_G} of the main text follows. 
As we see here, if $\mathcal{S}_{00}$ has a diverging value, it will enter the determinant of $\mathcal{S}$ and therefore force the $(\mathcal{S}^{-1})_{00}$ component to vanish in complete analogy to the simpler homogeneous case we studied in Sec.\,\ref{sec:holographicFS}. 
Given that using the harmonic sources we can evaluate a large enough sector of the $\mathcal{S}$-matrix, we can invert it approximately and obtain a required value of $(\mathcal{S}^{-1})_{00}$. However this treatment is not feasible in practice. The other way of obtaining  the alternative quantization picture is by directly setting the boundary condition for $b(x) = 1$ and read out the profile of $a(x)$. In this way we directly measure the  $\tilde{\mathcal{S}}_{00}$ component and no matrix inversion is needed. This approach is equivalent to setting the bulk fermion mass to its negative value \eqref{eq:boundary_expansion}, as we discussed above and it is much less demanding, therefore we predominantly use it in this work. Nonetheless,  the rescaling of the fermionic wave function discussed in Appendix\,\ref{app:fermi_numerics} is different in this case. Consequently, the equations differ from the direct quantization ones. We find it important to check whether the two approaches do indeed lead to the equivalent results. On Fig.\,\ref{fig:altQ_two_treatments} we show that indeed the result obtained from inverting the $\cal{S}$-matrix for a set of 34 harmonic sources (left panel) coincides with the one which we get by  changing the sign of the bulk fermion mass (right panel).

One extra comment on the structure of the $\mathcal{S}$-matrix is in order. This clarifies the relation between the shifts in parameter $k$  and the values of the spectral function in the different Brillouin zones. 
It is useful to recall that the $k$-parameter is a part of the definition of the Bloch wave-function \eqref{equ:holographic_bloch}. The actual solution to the equations of motion near the boundary is
\begin{equation}
\label{equ:eikx_factors}
\psi(x,z) = e^{i k x} \left[a(x) (1 + \dots) + b(x) z^\alpha(1+ \dots)\right].
\end{equation}
From this solution we infer that $\vec{b} = \mathcal{S}[k] \cdot \vec{a}$. However, one can represent the same wave-function in a different way
\begin{gather}
\label{equ:eikx_p_factors}
\psi(x,z) = e^{i (k - p) x} \left[\tilde{a}(x) (1 + \dots) + \tilde{b}(x) z^\alpha(1+ \dots)\right], 
\\
\label{equ:ab_tilde}
\tilde{a}(x) \equiv e^{i p x} a(x), \qquad \tilde{b}(x) \equiv e^{i p x} b(x).
\end{gather}
The $\tilde{a}(x)$ and $\tilde{b}(x)$ functions are still periodic in the unit cell, therefore  \eqref{equ:eikx_p_factors} is a perfectly valid Bloch wave representation as well. However, the Bloch momentum is now different and the relation between $\tilde{b}(x)$ and $\tilde{a}(x)$ reads
\begin{equation}
\label{equ:Skp}
\vec{\tilde{b}} = \mathcal{S}[k - p] \cdot \vec{\tilde{a}}
\end{equation}
Recalling the definition \eqref{equ:ab_tilde} we can relate the Fourier component of $a(x)$ and $\tilde{a}(x)$:
\begin{equation}
a_l = \tilde{a}_{l+1}, \qquad b_n = \tilde{b}_{n+1},
\end{equation}
and together with \eqref{equ:Skp} it allows us to relate the components of $\mathcal{S}[k]$ from the different Brillouin zones
\begin{equation}
\mathcal{S}[k-p]_{nl} = \mathcal{S}[k]_{n-1,l-1}.
\end{equation}
This identity serves as another useful check of our numerical procedures and our results are in agreement with it.

\bibliographystyle{JHEP-2}
\bibliography{lattice_fermions}

\end{document}